\documentclass{article}

    \PassOptionsToPackage{numbers, compress}{natbib}

\usepackage[final,main]{neurips_2026}

\usepackage[utf8]{inputenc} 
\usepackage[T1]{fontenc}    
\usepackage{hyperref}       
\usepackage{url}            
\usepackage{booktabs}       
\usepackage{amsmath}        
\usepackage{amsfonts}       
\usepackage{amssymb}        
\usepackage{nicefrac}       
\usepackage{microtype}      
\usepackage{xcolor}         
\usepackage{graphicx}       
\usepackage{multirow}       
\usepackage{pifont}         
\usepackage{algorithm}      
\usepackage{algpseudocode}  
\usepackage{pgfplots}       
\pgfplotsset{compat=1.18}
\usepackage{subcaption}     
\usepackage{tabularx}       
\usepackage{array}          
\usepackage{needspace}
\usepackage{wrapfig}        

\definecolor{mygreen}{RGB}{34, 139, 34}
\definecolor{myred}{RGB}{200, 50, 50}
\newcommand{\cmark}{{\color{mygreen}\ding{51}}}
\newcommand{\xmark}{{\color{myred}\ding{55}}}

\title{Efficient Streaming Audio-Visual Target Speaker Extraction for Real-World Acoustic Scenes}

\author{%
  {\normalfont %
    \textbf{Wendi Sang}\textsuperscript{1,2,3,}\thanks{Equal contribution.},\quad
    \textbf{Kai Li}\textsuperscript{2,}\footnotemark[1],\quad
    \textbf{Yifan Li}\textsuperscript{1,3}
  } \\
  {\normalfont %
    \textbf{Jianqiang Huang}\textsuperscript{1,3,}\thanks{Corresponding author.},\quad
    \textbf{Xiaolin Hu}\textsuperscript{2,4,}\footnotemark[2]
  } \\[2pt]
  {\normalfont \textsuperscript{1}School of Computer Technology and Application, Qinghai University,}\\
  {\normalfont Xining 810016, China}\\
  {\normalfont \textsuperscript{2}Department of Computer Science and Technology, Institute for AI, BNRist,}\\
  {\normalfont Tsinghua Laboratory of Brain and Intelligence (THBI),}\\
  {\normalfont IDG/McGovern Institute for Brain Research, Tsinghua University, Beijing 100084, China}\\
  {\normalfont \textsuperscript{3}Intelligent Computing and Application Laboratory of Qinghai Province,}\\
  {\normalfont Qinghai University, Xining 810016, China}\\
  {\normalfont \textsuperscript{4}Chinese Institute for Brain Research (CIBR), Beijing 100010, China}
}

\begin{document}
\raggedbottom

\maketitle

\begin{abstract}
Streaming audio-visual target speaker extraction (AVTSE) is essential for latency-sensitive applications. However, current causal systems face two coupled gaps. First, standard benchmarks fix the number of speakers at two and ignore the near-/far-field interference, reverberation, and sudden noise of real scenes. Second, high-performance causal architectures are too heavy for edge deployment. Lightweight alternatives try to reclaim capacity by repeatedly invoking a small separator, which offsets the savings. To address the first gap, we release RealSSA, a realistic AVTSE benchmark consisting of RealSSA-Sim and RealSSA-Real. RealSSA-Sim provides controllable simulated scenes with 1--5 dynamic near-field interfering speakers; scenes with far-field interference additionally include 10--15 far-field interfering speakers. RealSSA-Real provides a held-out real-recorded evaluation set. To address the second gap, we propose Falcon, a causal AVTSE method. Its Separator performs multi-scale time-frequency separation in a single encoder--decoder pass. We train Falcon with an encoder-space speaker contrastive loss. This loss suppresses near-field leakage at zero inference cost and transfers across backbones. Falcon achieves state-of-the-art extraction quality on RealSSA-Sim, LRS2-2Mix, LRS3-2Mix, and VoxCeleb2-2Mix. Compared with the causal AV-TFGridNet baseline, it cuts parameters by 91.9\%, computation by 20.2$\times$, and GPU inference latency by 11.1$\times$. Code and demo are available at \url{https://cslikai.cn/Falcon/}.
\end{abstract}

\section{Introduction}

Audio-visual target speaker extraction (AVTSE) leverages a speaker's visual stream, typically lip movements, as enrollment cues to separate the target speaker's voice from a multi-speaker mixture~\cite{li2025advances,ephrat2018looking}. Although this setting has often been described in prior work as audio-visual speech separation~\cite{gao2021visualvoice,lee2021looking,li2024iianet,li2024audio}, these methods all condition on the visual stream of a single target speaker, which by definition is target speaker extraction; therefore, we use the term AVTSE throughout this paper.
Many latency-sensitive applications now demand \emph{causal} AVTSE, such as video conferencing~\cite{gruttadauria2024online} and hearing-assistive devices~\cite{dhanjal2019tools}. However, the current causal AVTSE methods fall short along two axes: the data used to train and evaluate causal models, and the architecture of the causal model itself.


\begin{figure}[h]
  \centering
  \begin{subfigure}[b]{0.18\textwidth}
    \includegraphics[width=\textwidth,page=1]{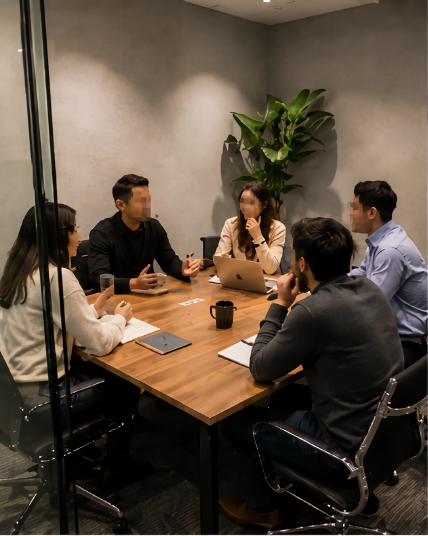}
    \caption{Small Rooms}
  \end{subfigure}
  \hspace{0.01\textwidth}
  \begin{subfigure}[b]{0.18\textwidth}
    \includegraphics[width=\textwidth,page=2]{figures/scene_panels.pdf}
    \caption{Coffee Shops}
  \end{subfigure}
  \hspace{0.01\textwidth}
  \begin{subfigure}[b]{0.18\textwidth}
    \includegraphics[width=\textwidth,page=3]{figures/scene_panels.pdf}
    \caption{Shopping Malls}
  \end{subfigure}
  \hspace{0.01\textwidth}
  \begin{subfigure}[b]{0.18\textwidth}
    \includegraphics[width=\textwidth,page=4]{figures/scene_panels.pdf}
    \caption{Outdoor Streets}
  \end{subfigure}

  \caption{Real-world AVTSE scenarios motivating RealSSA. The four scene types, ranging from small rooms to outdoor streets, cover varying number of speakers, near-/far-field interference, reverberation, and sudden environmental noise beyond standard fixed-speaker benchmarks.}
  \label{fig:Real_world_scenarios}
  \vspace{-8pt}
\end{figure}

Existing AVTSE benchmarks do not reflect the acoustic conditions of deployment. Some benchmarks built on LRS2 \cite{afouras2018deep}, LRS3 \cite{afouras2018lrs3}, and VoxCeleb2 \cite{chung2018voxceleb2} fix the number of speakers at two, draw mixtures at controlled SNRs, and hold room acoustics and background noise constant, leaving out the near-/far-field geometry, reverberation, and sudden noise events that characterize everyday recordings (Figure~\ref{fig:Real_world_scenarios}) \cite{wichern2019wham,yang2024realman}. None of them are paired with a real-recorded evaluation set, and reported gains offer little evidence that a causal model will hold up once deployed. Therefore, the AVTSE task requires a benchmark that can simulate simultaneous changes in the number of speakers, spatial geometry, and noise, while also providing real recordings to evaluate real-world deployment performance.

On the model side, current causal AVTSE faces an efficiency gap. Causal variants of strong non-causal architectures preserve the extraction quality. Some methods include AV-TFGridNet~\cite{pan2023scenario}, AV-Mossformer2~\cite{zhao2025clearervoice}, and RTFSNet~\cite{pegg2024rtfsnet}, obtained under the protocol of SwiftNet~\cite{Sang2025AFA}. However, their parameter counts and per-frame FLOPs push inference latency beyond real-time budgets on edge hardware. Lightweight alternatives such as SwiftNet~\cite{Sang2025AFA} recover capacity by repeatedly invoking a small separator after audio-visual fusion. Yet each additional cycle multiplies per-frame computation and erodes the latency savings of the smaller module. On the training side, SI-SNR penalizes deviation from the clean target \cite{le2019sdr}, but does not explicitly push the estimate away from near-field speakers. This leaves residual leakage from spatially close talkers, and the leakage grows with the number of simultaneously active near-field speakers in realistic scenes~\cite{li2025advances,liu2024improving}.

To close this data gap, we release \textit{RealSSA}, a realistic AVTSE benchmark consisting of RealSSA-Sim and RealSSA-Real. RealSSA-Sim provides controllable simulated scenes with 1--5 dynamic near-field interfering speakers across four scene types; scenes with far-field interference additionally include 10--15 far-field interfering speakers. RealSSA-Real provides a held-out real-recorded evaluation set. We jointly sample room reverberation, ambient noise, and sudden noise events, so that the number of interfering speakers, spatial geometry, and environmental acoustics vary together rather than one axis at a time. RealSSA-Real further allows causal models trained on simulation to be evaluated on deployment-side distributions without adaptation.

To close the model-side gap, we propose \textit{Falcon}, a single-pass causal AVTSE method. At its core, the \textit{Separator} applies an encoder--decoder trunk to the audio-visual fused feature. In a single decoder traversal, it performs frequency-axis, time-axis, and joint time-frequency modeling, assigned to different decoder levels from coarse to fine scales. We further introduce an \textit{encoder-space speaker contrastive loss}, computed in the separator's STFT encoder space against the clean target and each near-field interferer. This loss suppresses residual leakage at zero inference cost and transfers across AVTSE backbones. On RealSSA-Sim, LRS2-2Mix~\cite{afouras2018deep}, LRS3-2Mix~\cite{afouras2018lrs3}, and VoxCeleb2-2Mix~\cite{chung2018voxceleb2}, Falcon matched or exceeded the strongest causal baselines~\cite{pan2023scenario,zhao2025clearervoice,Sang2025AFA,li2026efficient} in separation quality (Table~\ref{tab:main-results}). On RealSSA-Real, it achieved competitive perceptual quality, WER, and target-speaker similarity. Meanwhile, it reduced parameters, computational cost, and GPU inference latency by 91.9\%, 20.2$\times$, and 11.1$\times$ relative to the causal AV-TFGridNet \cite{pan2023scenario}.

\section{Related Works}

\subsection{AVTSE Benchmarks}
\label{subsec:tse_benchmarks}

AVTSE benchmarks can be organized along three axes: speaker configuration, acoustic conditions, and evaluation realism (Table~\ref{tab:avss_benchmarks}). Along the first axis, LRS2-2Mix \cite{afouras2018deep}, LRS3-2Mix \cite{afouras2018lrs3}, and VoxCeleb2-2Mix \cite{chung2018voxceleb2} fix two near-field speakers per mixture, and therefore do not expose models to varying counts of near- and far-field interferers. These three corpora are used by most AVTSE models \cite{lee2021looking,li2024audio,li2024iianet,pegg2024rtfsnet,li2026efficient} and their causal variants \cite{Sang2025AFA}. Along the second axis, these corpora inherit acoustic conditions from the source videos, and offer no controlled reverberation or sudden noise events. The LRS3+WHAM! noisy protocol, reused by \cite{martel2023audio,xue2025coarse,zhang2025audio}, adds WHAM! \cite{wichern2019wham} background noise at SNR roughly in $[-6,+3]$~dB, but still keeps two speakers, anechoic mixing, and no sudden noise. The NTCD-TIMIT-based 2-speaker+noise protocol \cite{martel2023audio,abdelaziz2017ntcdtimit} likewise fixes two speakers, and varies only the noise SNR in $[-5,20]$~dB. Along the third axis, evaluation relies on held-out splits of the same source videos rather than independent real recordings. The gap between simulated mixtures and true on-device capture therefore remains undiagnosed. As summarized in Table~\ref{tab:avss_benchmarks}, RealSSA is the only AVTSE benchmark that satisfies all three axes jointly.

\begin{table}[h]
\centering
\caption{AVTSE benchmarks organized along three axes: speaker configuration, acoustic conditions, and evaluation realism. ``NF'' and ``FF'' denote near- and far-field sources; ``Env.~noise'' means an external noise corpus is mixed in.}
\label{tab:avss_benchmarks}
\resizebox{\linewidth}{!}{
\begin{tabular}{lcccccc}
\toprule
Benchmark & NF speakers & FF speakers & Reverb control & Env. noise & Sudden noise & Real-recorded eval \\
\midrule
NTCD-TIMIT+noise [\citeyear{abdelaziz2017ntcdtimit}] & 2 & \xmark & \xmark & \cmark & \xmark & \xmark \\
LRS2-2Mix [\citeyear{afouras2018deep}]          & 2 & \xmark & \xmark & \xmark & \xmark & \xmark \\
LRS3-2Mix [\citeyear{afouras2018lrs3}]          & 2 & \xmark & \xmark & \xmark & \xmark & \xmark \\
VoxCeleb2-2Mix [\citeyear{chung2018voxceleb2}]  & 2 & \xmark & \xmark & \xmark & \xmark & \xmark \\
LRS3+WHAM! [\citeyear{martel2023audio}] & 2 & \xmark & \xmark & \cmark & \xmark & \xmark \\
\textbf{RealSSA (ours)} & \textbf{2--6} & \textbf{\cmark} & \textbf{\cmark} & \textbf{\cmark} & \textbf{\cmark} & \textbf{\cmark} \\
\bottomrule
\end{tabular}}
  \vspace{-8pt}
\end{table}

\subsection{Causal Audio-Visual Target Speaker Extraction}
\label{subsec:causal_tse}

Causal AVTSE methods mainly differ in two aspects: separator architecture and training objective. On the architecture side, causal evaluation largely follows the protocol of SwiftNet \cite{Sang2025AFA}, which provides a recipe for converting non-causal AVTSE backbones into their causal counterparts by restricting temporal operations to past and current frames; earlier time-domain separators \cite{luo2019conv,tao2025seanet,lin2023av,li2024audio} and recent time-frequency separators are causalized under this protocol. Among the time-frequency family, AV-TFGridNet \cite{pan2023scenario} and AV-Mossformer2 \cite{zhao2025clearervoice} retain strong extraction quality at the cost of large parameter counts and heavy per-frame computation, while SwiftNet and the causalized RTFSNet \cite{pegg2024rtfsnet} and AVLiT \cite{martel2023audio} variants reuse a single compact block across multiple iterative cycles, trading parameters for per-frame FLOPs. Falcon instead employs level-specific time-frequency blocks within a single encoder--decoder forward pass.

On the training side, AVTSE systems were typically optimized with waveform reconstruction losses such as SI-SNR \cite{le2019sdr}, and prior attempts to suppress residual interferers either adjust the training mixture distribution or add routing-level gates over the separator output \cite{liu2024improving,pan2023scenario} rather than explicitly separating the estimate from each interfering speaker. Falcon instead adds an encoder-space speaker contrastive loss, discarded at inference time.

\section{The RealSSA Dataset}
\label{sec:srav}

RealSSA consists of two parts: RealSSA-Sim, a controllable simulated-scene suite used for training, validation, and testing, and RealSSA-Real, a held-out real-recorded evaluation set used to assess generalization under non-simulated acoustic conditions. RealSSA-Sim provides systematic control over speaker configuration, source distance, reverberation, environmental noise, and sudden noise events, while RealSSA-Real reflects raw on-device recordings in everyday acoustic environments.

\textbf{RealSSA-Sim: Simulated Scene Suite.}
RealSSA-Sim defines four acoustic scene types that span a wide range of room sizes, reverberation levels, and noise conditions (Table~\ref{tab:srav_config}). Small meeting rooms have short RT60 and only near-field interferers, representing close-range conversations. Medium indoor spaces introduce far-field speakers and moderate reverberation. Large public spaces further increase RT60 and the number of interferers, reflecting scenarios such as lecture halls. Outdoor environments have minimal reverberation but high ambient noise levels. Across the four scenes, the number of near-field interfering speakers varies from 1 to 5; scenes with far-field interference additionally include 10--15 far-field interfering speakers. Each scene independently samples RT60, near-/far-field geometry, environmental noise, and sudden noise events. We classify an interferer as \emph{near-field} when it lies within $3\,\mathrm{m}$ of the microphone and as \emph{far-field} otherwise; this threshold approximates the critical distance in typical indoor rooms, beyond which reverberant energy dominates the direct path~\cite{kuttruff2016room}. Per-scene spatial layouts are illustrated in Figure~\ref{fig:scene_layouts} (Appendix~\ref{app:realssa:source}).

\begin{table}[h]
\centering
\caption{Acoustic scene configurations in RealSSA-Sim, covering four room types with controlled reverberation, speaker distance, and noise conditions. Near speakers and Far speakers denote interfering speakers and exclude the target.}
\label{tab:srav_config}
\resizebox{\linewidth}{!}{
\begin{tabular}{lcccccc}
\toprule
Scene & Space size & RT60 & Near speakers & Far speakers & Env. noise SIR & Sudden noise \\
\midrule
Small meeting
& \(4\text{-}6 \times 3\text{-}5 \times 2.5\text{-}3\,\mathrm{m}\)
& 0.3--0.6 s
& 1--5, 0.5--1.5 m
& none
& 5--15 dB
& 30\%, 1 event \\
Medium indoor
& \(8\text{-}15 \times 8\text{-}12 \times 3\text{-}4\,\mathrm{m}\)
& 0.5--0.9 s
& 1--3, 1.0--2.5 m
& 10--15, 3--15 m
& 5--10 dB
& 40\%, 1--2 events \\
Large public
& \(20\text{-}40 \times 15\text{-}30 \times 4\text{-}8\,\mathrm{m}\)
& 1.0--2.0 s
& 1--4, 1.0--3.0 m
& 10--15, 5--20 m
& 2--8 dB
& 50\%, 1--3 events \\
Outdoor
& \(50 \times 50 \times 20\,\mathrm{m}\)
& 0.05--0.1 s
& 1--4, 1.0--4.0 m
& 10--15, 5--25 m
& $-2$--$5$ dB
& 40\%, 1--3 events \\
\bottomrule
\end{tabular}}
  \vspace{-8pt}
\end{table}

Target audio-visual data are sampled from LRS3 \cite{afouras2018lrs3}, while near-field and far-field interfering speakers come from LibriSpeech \cite{panayotov2015librispeech}. Non-speech noise is drawn from WHAM! \cite{wichern2019wham} for continuous backgrounds and Hive \cite{li2026semantically} for sudden noise events. 
The generation pipeline proceeds in three stages. \textbf{(1) Room simulation.} For each sample, we randomly select a scene type and sample the room geometry and RT60 from the corresponding ranges in Table~\ref{tab:srav_config}. Room impulse responses are computed with gpuRIR \cite{diaz2021gpurir}. \textbf{(2) Source placement.} The target speaker is placed at a random position within the near-field range of the selected scene. Interfering speakers are placed according to the scene-specific near-field and far-field distance ranges. \textbf{(3) Mixing.} The target speech is normalized to $-23$ LUFS following the EBU R128 broadcast loudness standard \cite{ebu2020r128}. Interfering speakers and environmental noise are added at scene-specific SIR ranges, and the final mixture is peak-limited to $-1\,\mathrm{dBFS}$. All audio is single-channel at $16\,\mathrm{kHz}$. 
RealSSA-Sim contains $100{,}826$ training, $9{,}133$ validation, and $9{,}169$ test samples, totaling approximately $153$ hours. The four scene types are uniformly sampled, with an average of $2.5$ near-field interfering speakers per sample. Source corpora and scene-design rationale are detailed in Appendix~\ref{app:realssa:source}, the full mixing pipeline in Appendix~\ref{app:realssa:pipeline}, and per-split distribution statistics in Appendix~\ref{app:realssa:stats}. We evaluated the generalization from RealSSA-Sim training to RealSSA-Real in Section~\ref{sec:generalization}.

\textbf{RealSSA-Real: Real-World Evaluation Set.} In addition to the simulated suite, we collect a real-recorded evaluation set, denoted as RealSSA-Real, to assess generalization under non-simulated acoustic conditions. RealSSA-Real covers everyday indoor, public, and outdoor environments with natural background noise, room reverberation, and uncontrolled interfering speech. It is recorded with a single mobile device and preserves the raw acoustic conditions of real on-device capture. The recordings are post-processed only to match the input format used by the AVTSE pipeline, including audio resampling to 16 kHz, video conversion to 25 fps, and extraction of 96$\times$96 grayscale mouth regions.
RealSSA-Real is used exclusively for evaluation and is never used for training or validation. We assess perceptual quality with DNSMOS and NISQA, speech recognizability with WER, and target-speaker consistency with speaker similarity. Detailed recording protocol, environment composition, and privacy processing are provided in Appendix~\ref{app:realssa-real}.

\section{Methods}

\subsection{Overall Pipeline}
\label{subsec:overall_pipeline}

\begin{figure}[h]
  \centering
  \includegraphics[width=\textwidth]{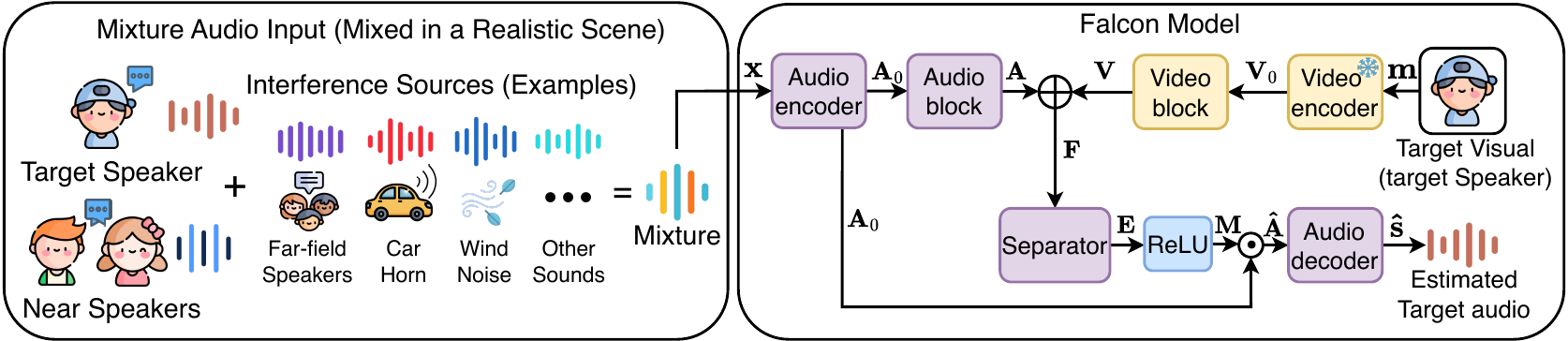}
  \caption{Overall pipeline of Falcon for audio-visual target speaker extraction.}
  \label{fig:overall_pipeline}
  \vspace{-1pt}
\end{figure}

Given a mixture waveform $\mathbf{x}\in\mathbb{R}^{L}$ of length $L$ and a mouth video $\mathbf{m}\in\mathbb{R}^{H\times W\times T_{\text{v}}}$ of the target speaker with spatial resolution $H\times W$ and $T_{\text{v}}$ frames, Falcon aims to extract the target speech $\mathbf{s}\in\mathbb{R}^{L}$ as captured at the microphone. The mixture is modeled as
$
\mathbf{x}=\mathbf{s} + \sum_{k=1}^{K}\mathbf{b}_k + \mathbf{f} + \mathbf{n},
$
where $\mathbf{s},\mathbf{b}_k,\mathbf{f}\in\mathbb{R}^{L}$ denote the target speech, the $k$-th near-field interfering speech signal, and the sum of far-field interfering speech signals, respectively, all captured at the microphone, including room reverberation. The term $\mathbf{n}\in\mathbb{R}^{L}$ denotes additive environmental noise, and $\mathbf{f}=\mathbf{0}$ for scenes without far-field interferers.

As illustrated in Figure~\ref{fig:overall_pipeline}, the FRCNN visual backbone used in CTCNet~\cite{li2024audio}, frozen during training, extracts mouth features from $\mathbf{m}$ and maps them through a visual bottleneck to obtain $\mathbf{V}_0\in\mathbb{R}^{C_{\text{v}}\times T_{\text{v}}}$, where $C_{\text{v}}$ is the visual feature dimension. The audio encoder $E_{\text{a}}:\mathbb{R}^{L}\to\mathbb{R}^{C_{\text{e}}\times T\times N_{\text{f}}}$ applies STFT to its input, concatenates the real part, imaginary part, and magnitude spectrum as a three-channel tensor, and processes them through a 2D convolutional layer; we write $\mathbf{A}_0=E_{\text{a}}(\mathbf{x})$, where $T$ and $N_{\text{f}}$ denote the number of STFT frames and frequency bins, and $C_{\text{e}}$ is the audio encoder output dimension. An audio bottleneck then projects $\mathbf{A}_0$ to $\mathbf{A}\in\mathbb{R}^{C\times T\times N_{\text{f}}}$.

We refine each modality before fusion. The audio feature $\mathbf{A}$ is processed by an AudioBlock $\mathcal{B}_{\mathrm{a}}$, whose structure is the same as the separator detailed in Section \ref{sec:separator}, yielding $\tilde{\mathbf{A}}=\mathcal{B}_{\mathrm{a}}(\mathbf{A})\in\mathbb{R}^{C\times T\times N_{\mathrm{f}}}$. The visual feature $\mathbf{V}_0$ is processed by the VideoBlock $\mathcal{B}_{\mathrm{v}}$ of SwiftNet~\cite{Sang2025AFA}, which aggregates temporal context with a causal LSTM and upsamples to the audio frame rate, yielding $\tilde{\mathbf{V}}=\mathcal{B}_{\mathrm{v}}(\mathbf{V}_0)\in\mathbb{R}^{C\times T}$. $\tilde{\mathbf{V}}$ is expanded along the frequency axis and added to $\tilde{\mathbf{A}}$, producing $\mathbf{F}\in\mathbb{R}^{C\times T\times N_{\mathrm{f}}}$. Then, the fused feature $\mathbf{F}$ is processed by the separator $\mathcal{B}_{\mathrm{a}}$ to obtain the target-speaker feature $\mathbf{E}\in\mathbb{R}^{C\times T\times N_{\mathrm{f}}}$. Each application of $\mathcal{B}_{\mathrm{a}}$ completes multi-scale time-frequency modeling in one encoder--decoder forward pass and is not unrolled along the time axis.
Then, the mask generator predicts a time-frequency mask $\mathbf{M}\in\mathbb{R}^{C_{\text{e}}\times T\times N_{\text{f}}}$ from $\mathbf{E}$. We split the channel dimensions of $\mathbf{M}$ and $\mathbf{A}_0$ evenly into real and imaginary sub-bands, denoted $\mathbf{M}=[\mathbf{M}_{\text{r}};\mathbf{M}_{\text{i}}]$ and $\mathbf{A}_0=[\mathbf{A}_{0,\text{r}};\mathbf{A}_{0,\text{i}}]$, and obtain the masked embedding by complex multiplication,
\begin{equation}
\hat{\mathbf{A}}=
\begin{bmatrix}
\mathbf{M}_{\text{r}}\odot\mathbf{A}_{0,\text{r}}-\mathbf{M}_{\text{i}}\odot\mathbf{A}_{0,\text{i}}\\
\mathbf{M}_{\text{r}}\odot\mathbf{A}_{0,\text{i}}+\mathbf{M}_{\text{i}}\odot\mathbf{A}_{0,\text{r}}
\end{bmatrix}
\in\mathbb{R}^{C_{\text{e}}\times T\times N_{\text{f}}},
\end{equation}
where $\odot$ is element-wise multiplication. The audio decoder finally maps $\hat{\mathbf{A}}$ back to real and imaginary spectral components and reconstructs the target speech $\hat{\mathbf{s}}\in \mathbb{R}^{L}$ via iSTFT. 

Separator plays a central role in Falcon, since it is responsible for extracting the $\mathbf{F}$ into the target speaker features $\mathbf{E}$. We next introduce our single-pass multi-scale design for Separator.

\subsection{Single-Pass Multi-Scale Separator}
\label{sec:separator}

\begin{figure}[t]
  \centering
  \includegraphics[width=0.85\textwidth]
  {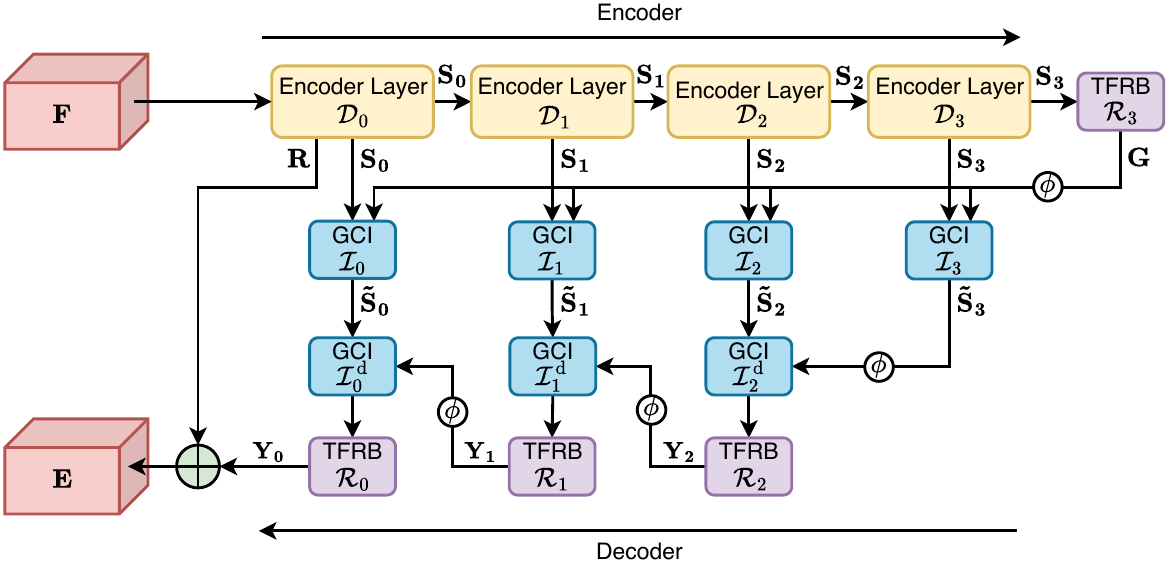}
  \caption{Single-Pass Multi-Scale Separator. $\phi$ denotes the upsampling operation.}
  \label{fig:separator}
  \vspace{-8pt}
\end{figure}

The Separator is an asymmetric U-Net (Figure~\ref{fig:separator}): time-frequency modeling is concentrated at the encoder bottleneck to form a single coarse scene context, while a level-specific TF block at every decoder resolution recovers fine spectral structure step by step. A single encoder--decoder pass therefore already yields four distinct TF scales, replacing the multi-iteration recurrence of SwiftNet~\cite{Sang2025AFA}.

The input feature $\mathbf{F}$ first passes through a depthwise $1\times1$ convolution with PReLU to form a residual branch $\mathbf{R}\in\mathbb{R}^{C\times T\times N_{\text{f}}}$, which serves both as the entry to the hidden path below and as a global skip to the separator output. A pointwise $1\times1$ convolution with PReLU then projects $\mathbf{R}$ to the hidden channel dimension $C_{\text{h}}$, and a streaming-causal layer normalization $\operatorname{LN}_{\text{c}}$ produces $\mathbf{X}_0\in\mathbb{R}^{C_{\text{h}}\times T\times N_{\text{f}}}$. For any input $\mathbf{Z}$, $\operatorname{LN}_{\text{c}}$ is defined per time-frequency position as
\begin{equation}
[\operatorname{LN}_{\text{c}}(\mathbf{Z})]_{c,t,f}
= \gamma_{c}\cdot\frac{Z_{c,t,f}-\mu_{t,f}}{\sqrt{\sigma^{2}_{t,f}+\epsilon}}+\beta_{c},\qquad
\mu_{t,f}=\tfrac{1}{C_{\text{h}}}\!\sum_{c} Z_{c,t,f},
\end{equation}
with $\sigma^{2}_{t,f}$ defined analogously, $\gamma,\beta\in\mathbb{R}^{C_{\text{h}}}$ learnable, and a small $\epsilon$ for stability. $\operatorname{LN}_{\text{c}}$ is frame-local and cache-free, and is reused throughout the blocks below.

\begin{figure}[t]
  \centering
  \includegraphics[width=0.9\textwidth]
  {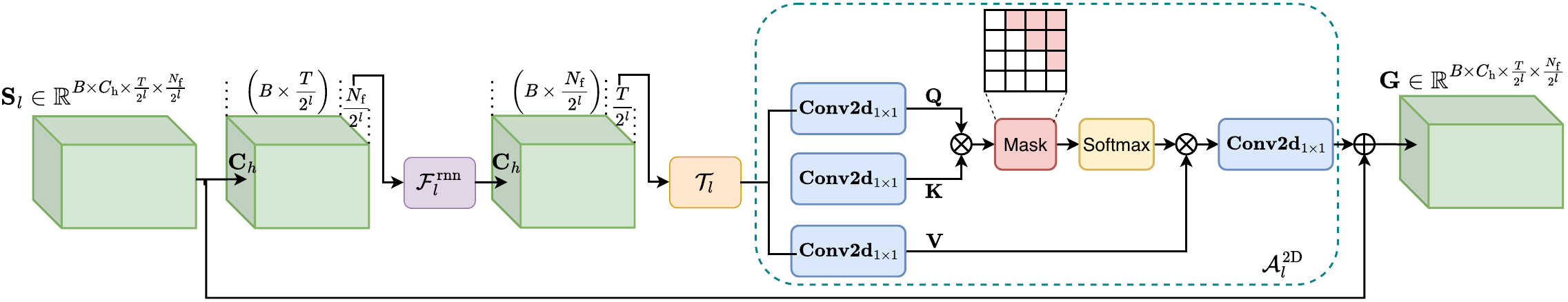}
  \caption{Time-Frequency Refinement Block.}
  \label{fig:TFRB}
  \vspace{-8pt}
\end{figure}

\textbf{Encoder.}
The encoder has $L{=}4$ levels. $\mathcal{D}_0$ is a stride-$1$ depthwise $3{\times}3$ convolution; $\mathcal{D}_1,\mathcal{D}_2,\mathcal{D}_3$ are stride-$2$ depthwise $5{\times}5$ convolutions that halve both axes. All convolutions use left-aligned padding along time and symmetric padding along frequency, adding no temporal lookahead. Starting from $\mathbf{S}_0=\mathcal{D}_0(\mathbf{X}_0)$, subsequent levels are produced in order $l=1,2,3$ by $\mathbf{S}_l=\mathcal{D}_l(\mathbf{S}_{l-1})$, giving
\begin{equation}
\mathbf{S}_l \in \mathbb{R}^{C_{\text{h}}\times \frac{T}{2^l}\times \frac{N_{\text{f}}}{2^l}}, \quad l=0,\dots,3.
\end{equation}
Each $\mathbf{S}_l$ is kept as a skip. Time-frequency modeling is applied only at the bottleneck,
\begin{equation}
\mathbf{G}=\mathcal{R}_{3}(\mathbf{S}_{3}) \in \mathbb{R}^{C_{\text{h}}\times \frac{T}{8}\times \frac{N_{\text{f}}}{8}},
\end{equation}
where $\mathcal{R}_{l}(\cdot)$ is the Time-Frequency Refinement Block (TFRB) at level $l$ (Figure~\ref{fig:TFRB}); we identify $\mathbf{S}_3 \equiv \mathbf{G}$ in what follows.

Each TFRB is a cascade of three stages applied in order: first a bidirectional LSTM along frequency $\mathcal{F}_{l}^{\text{rnn}}$, then a unidirectional LSTM along time $\mathcal{T}_{l}$, and finally a 2D self-attention $\mathcal{A}_{l}^{\text{2D}}$,
\begin{equation}
\mathcal{R}_{l}(\mathbf{X})
=
\mathcal{A}_{l}^{\text{2D}}
\left(
\mathcal{T}_{l}
\left(
\mathcal{F}_{l}^{\text{rnn}}(\mathbf{X})
\right)
\right),
\end{equation}
where $\mathcal{A}_{l}^{\text{2D}}$ operates on the flattened time-frequency plane with a block-causal mask: a query at $(t,f)$ attends to every key $(t',f')$ with $t'\le t$. Each stage is wrapped in a pre-norm residual block with $\operatorname{LN}_{\text{c}}$.

\textbf{Global context injection (GCI) module.}
For $l=0,\dots,3$, we inject $\mathbf{G}$ into each skip via
\begin{equation}
\tilde{\mathbf{S}}_l=\mathcal{I}_l(\mathbf{S}_l,\mathbf{G})
=
\phi_{\text{loc}}(\mathbf{S}_l)
\odot
\sigma\!\left(\phi_{\text{gate}}(\mathbf{G})\right)
+
\phi_{\text{glb}}(\mathbf{G}),
\end{equation}
where $\mathbf{G}$ is first upsampled to the resolution of $\mathbf{S}_l$ by nearest-neighbor interpolation when needed, and $\phi_{\text{loc}},\phi_{\text{glb}},\phi_{\text{gate}}$ are depthwise convolutions with $\operatorname{LN}_{\text{c}}$. The gate $\sigma(\phi_{\text{gate}}(\mathbf{G}))$ lets the global context emphasize or suppress frequency bands of the local skip.

\textbf{Decoder.}
The decoder restores resolution coarse-to-fine: for $l=2,1,0$ in that order,
\begin{equation}
\mathbf{Y}_{l}
=
\mathcal{R}_{l}
\left(
\mathcal{I}^{\text{d}}_{l}(\tilde{\mathbf{S}}_{l},\mathbf{Y}_{l+1})
+
\mathbf{S}_{l}
\right),
\end{equation}
with $\mathbf{Y}_3 \equiv \tilde{\mathbf{S}}_3$ and $\mathcal{I}^{\text{d}}_{l}(\cdot)$ the decoder-side GCI module whose coarser argument is first upsampled. The $+\mathbf{S}_l$ term is an identity residual that preserves gradient flow and a raw spectral reference when the gate in $\mathcal{I}^{\text{d}}_l$ collapses $\tilde{\mathbf{S}}_l$ toward zero. Since each $\mathcal{R}_l$ operates at a different resolution, the four TFRBs span four TF scales in one forward pass.

The decoder output $\mathbf{Y}_0\in\mathbb{R}^{C_{\text{h}}\times T\times N_{\text{f}}}$ is projected to $C$ channels and added to the residual branch,
$
\mathbf{E}
=
\mathcal{O}(\mathbf{Y}_0)
+
\mathbf{R},
$
where $\mathcal{O}(\cdot)$ is a pointwise projection.

\subsection{Encoder-Space Speaker Contrastive Loss}

Standard reconstruction losses such as SI-SNR minimize the distance between the estimate and the clean target but do not explicitly push the estimate away from near-field interferers. We add a contrastive loss that reuses the audio encoder $E_{\text{a}}$ for both the query and the references, so no separate encoder is introduced.

The query is the estimated target in encoder space, $\hat{\mathbf{A}}\in\mathbb{R}^{C_{\text{e}}\times T\times N_{\text{f}}}$, obtained from the complex masking of Section~\ref{subsec:overall_pipeline}. The positive and negative references are the clean target $\mathbf{s}$ and each near-field interferer $\mathbf{b}_k$, both captured at the microphone and therefore carrying the same room response as the mixture, passed through $E_{\text{a}}$ with stop-gradient on the reference path:
\begin{equation}
\mathbf{A}^{+}=\operatorname{sg}(E_{\text{a}}(\mathbf{s})),\qquad
\mathbf{A}^{-}_{k}=\operatorname{sg}(E_{\text{a}}(\mathbf{b}_k)),\quad k=1,\dots,K.
\end{equation}
Gradients flow into $E_{\text{a}}$ only through the query $\hat{\mathbf{A}}$, not through the references. Far-field interferers are excluded from the negative set: both distance attenuation and, in reverberant scenes, late reverberation drive their residual below the SI-SNR error floor, so adding them as negatives yields no additional signal. A lightweight speaker pooling function $\mathcal{P}_{\text{spk}}(\cdot)$ then maps each encoder-space feature to a unit-norm speaker embedding by averaging over frequency, adaptively pooling over time, and projecting it to a fixed-dimensional vector:
\begin{equation}
\mathbf{q}=\mathcal{P}_{\text{spk}}(\hat{\mathbf{A}}),\qquad
\mathbf{p}=\mathcal{P}_{\text{spk}}(\mathbf{A}^{+}),\qquad
\mathbf{e}_k^{-}=\mathcal{P}_{\text{spk}}(\mathbf{A}^{-}_{k}).
\end{equation}

Writing $s_k\!\triangleq\!\cos(\mathbf{q},\mathbf{e}_k^{-})$, the contrastive loss combines a positive attraction term with a softmax-weighted, margin-based repulsion over negatives:
\begin{equation}
\mathcal{L}_{\text{spk}}
=
\bigl(1-\cos(\mathbf{q},\mathbf{p})\bigr)
+
\sum_{k=1}^{K}
\frac{\exp(\tau\,s_k)}{\sum_{j=1}^{K}\exp(\tau\,s_j)}\,
\operatorname{ReLU}(s_k-m),
\end{equation}
where $m{=}0.5$ (a $60^{\circ}$ angular margin) deactivates the repulsion once $\mathbf{q}$ is sufficiently far from $\mathbf{e}_k^{-}$, leaving SI-SNR to handle the remaining residual. The softmax weight acts via an \emph{inverse} temperature $\tau$: $\tau{=}0$ gives uniform weighting across negatives, and larger $\tau$ sharpens the distribution toward the hardest negative. We linearly anneal $\tau$ from $0$ to $1$ during training.

The final training objective is $\mathcal{L}=\mathcal{L}_{\text{recon}}+\lambda_{\text{spk}}\,\mathcal{L}_{\text{spk}}$ with $\mathcal{L}_{\text{recon}}=-\operatorname{SI\text{-}SNR}(\hat{\mathbf{s}},\mathbf{s})$~\cite{le2019sdr} and $\lambda_{\text{spk}}=1.0$. Since query and reference signals share the same audio encoder, this loss can be readily attached to any audio encoder AVTSE backbone with no additional architectural changes required.

\section{Experiment Configurations}

\subsection{Dataset Setup}

We evaluated on three public AVTSE benchmarks (LRS2~\cite{afouras2018deep}, LRS3~\cite{afouras2018lrs3}, VoxCeleb2~\cite{chung2018voxceleb2}), RealSSA-Sim (Section~\ref{sec:srav}) and RealSSA-Real (Appendix~\ref{app:realssa-real}). Mixtures on the three public benchmarks followed the protocol of prior works~\cite{li2024audio,Sang2025AFA}. All clips were $16\,\mathrm{kHz}$ audio paired with $96\times96$ grayscale mouth crops at $25$\,fps, and dataset splits followed the official partitions of each corpus. RealSSA-Real was used exclusively for evaluation. Further preprocessing details are given in Appendix~\ref{app:dataset_setup}.

\subsection{Model Configurations and Evaluation Metrics}
\label{sec:training}

The separator hidden dimension $C_{\text{h}}$ was set to $64$, and the full width assignment across $C_{\text{e}}$, $C$, $C_{\text{v}}$, and $\dim(\mathbf{q})$ is tabulated in Appendix~\ref{app:arch_widths}. The audio encoder $E_{\text{a}}$ used an STFT with window size $256$ and hop $128$ ($8$\,ms hop at $16\,\mathrm{kHz}$), and fed the real, imaginary, and magnitude components as a three-channel input, following Section~\ref{subsec:overall_pipeline}. For all models trained in this work, we used AdamW~\cite{loshchilov2017decoupled} (learning rate $1\times10^{-3}$, weight decay $0.01$), with the learning rate halved whenever the validation loss plateaued for $15$ consecutive epochs and early stopping applied after a patience of $30$ epochs. Full training hyperparameters are provided in Appendix~\ref{app:training_protocol}. We evaluated separation quality with SI-SNRi and SDRi~\cite{vincent2006performance} on the simulated benchmarks, and non-intrusive perceptual quality on RealSSA-Real with DNSMOS~\cite{reddy2022dnsmos} and NISQA~\cite{mittag2021nisqa}. On RealSSA-Sim and RealSSA-Real, we additionally measured word error rate (WER) with Mega-ASR~\cite{xie2026mega} and target-speaker similarity with a pretrained ResNet34 speaker encoder from WeSpeaker~\cite{wang2023wespeaker}. We also reported the number of parameters, MACs, CPU and GPU inference latency, and memory usage to assess computational efficiency, all measured on a $2$\,s, $16\,\mathrm{kHz}$ input clip paired with $50$ mouth-crop frames at $25$\,fps. Metric definitions and the inference-latency measurement protocol are detailed in Appendix~\ref{app:eval_protocol}.

\section{Results and Analysis}

\subsection{Main Results}

We retrained each baseline separately on RealSSA-Sim and VoxCeleb2-2Mix. Within each dataset, all models used the same training, validation, and test splits, preprocessing pipeline, optimizer, and learning-rate schedule. We then evaluated the models trained on RealSSA-Sim on RealSSA-Real. Dolphin~\cite{li2026efficient} and AV-Mossformer2~\cite{zhao2025clearervoice} were converted to causal variants (Appendix~\ref{app:baseline_causal}). Baseline results on LRS2-2Mix and LRS3-2Mix are from Swift-Net~\cite{Sang2025AFA}.

\textbf{Signal-level separation quality.} Table~\ref{tab:main-results} reports SI-SNRi and SDRi on the four simulated benchmarks. Falcon achieved the highest SI-SNRi and SDRi on all four. Its SI-SNRi gain over Swift-Net-12 was largest on VoxCeleb2-2Mix, reaching $1.1$\,dB. On RealSSA-Sim, which includes dynamic speaker counts, near-/far-field interference, and reverberation, Falcon achieved $7.7$\,dB SI-SNRi compared with $7.6$\,dB for AV-TF-GridNet. Appendix~\ref{app:scene_speaker_breakdown} provides results by scene type and near-field speaker count, and Appendix~\ref{app:qualitative} presents qualitative spectrogram comparisons.

\begin{table}[h]
  \caption{Performance of causal AVTSE methods. LRS2-2Mix and LRS3-2Mix baseline results are from Swift-Net; other baseline results use our retrained models. Bold and underline mark the best and second-best results.}
  \label{tab:main-results}
  \centering
  \resizebox{\textwidth}{!}{%
    \begin{tabular}{lccccccccc}
    \toprule
    \multirow{2}{*}{Method} & RealSSA-Real & \multicolumn{2}{c}{RealSSA-Sim} & \multicolumn{2}{c}{LRS2-2Mix} & \multicolumn{2}{c}{LRS3-2Mix} & \multicolumn{2}{c}{VoxCeleb2-2Mix} \\
    \cmidrule(lr){2-2} \cmidrule(lr){3-4} \cmidrule(lr){5-6} \cmidrule(lr){7-8} \cmidrule(lr){9-10}
     & DNSMOS $\uparrow$ & SI-SNRi $\uparrow$ & SDRi $\uparrow$ & SI-SNRi $\uparrow$ & SDRi $\uparrow$ & SI-SNRi $\uparrow$ & SDRi $\uparrow$ & SI-SNRi $\uparrow$ & SDRi $\uparrow$ \\
    \midrule
    AV-ConvTasNet~[\citeyear{zhang2021online}]   & 2.580 & 1.2  & 2.6 & 9.8  & 10.3 & 12.5 & 12.8 & 6.9  & 8.2  \\
    AV-Sepformer~[\citeyear{lin2023av}]          & 2.813 & 5.5  & 6.3 & 10.6 & 11.1 & 13.4 & 13.7 & 6.9  & 8.3  \\
    AV-TF-GridNet~[\citeyear{pan2023scenario}]   & 3.030 & \underline{7.6}  & \underline{8.1} & 12.7 & 13.1 & 14.3 & 14.8 & 5.8  & 7.7  \\
    RTFSNet-12~[\citeyear{pegg2024rtfsnet}]      & \underline{3.066} & 4.9  & 5.7 & 12.4 & 12.8 & 14.0 & 14.4 & 8.4  & 10.7 \\
    CTCNet~[\citeyear{li2024audio}]              & 2.840 & 5.4  & 6.2 & 8.0  & 8.8  & 9.4  & 10.2 & 2.5  & 4.7  \\
    AV-DPRNN~[\citeyear{tao2025seanet}]          & 2.901 & 5.4  & 6.2 & 10.2 & 10.7 & 11.9 & 12.3 & 7.8  & 9.0  \\
    AV-Mossformer2~[\citeyear{zhao2025clearervoice}] & 2.912 & 6.0  & 6.7 & 13.2 & 13.9 & \underline{16.2} & \underline{16.6} & 7.7  & 9.1  \\
    Swift-Net-12~[\citeyear{Sang2025AFA}]        & \textbf{3.276} & 7.3  & 7.8 & 13.9 & 14.1 & 16.0 & 16.3 & \underline{9.4}  & 10.8 \\
    Dolphin~[\citeyear{li2026efficient}]         & 2.960 & 6.7  & 7.3 & \underline{14.0} & \underline{14.6} & 14.3 & 14.9 & 9.2  & \underline{11.1} \\
    \midrule
    \textbf{Falcon (Ours)} & \textbf{3.276} & \textbf{7.7} & \textbf{8.2} & \textbf{14.1} & \textbf{14.7} & \textbf{16.4} & \textbf{16.7} & \textbf{10.5} & \textbf{11.8} \\
    \bottomrule
  \end{tabular}%
  }
  \vspace{-8pt}
\end{table}

\textbf{Speech quality and target-speaker consistency.} On RealSSA-Real, Falcon achieved the highest DNSMOS of $3.276$, tying Swift-Net-12 and exceeding AV-TF-GridNet ($3.030$) and Dolphin ($2.960$) (Table~\ref{tab:main-results}). Table~\ref{tab:wer-speaker-similarity} complements this perceptual evaluation with WER and target-speaker similarity. Among extraction systems, Falcon achieved the lowest RealSSA-Sim WER at $25.47\%$, improving on AV-TF-GridNet by $1.84$ percentage points. Its RealSSA-Real WER of $32.74\%$ ranked second, $0.74$ points behind AV-Mossformer2. Falcon also achieved the highest mean target-speaker similarity on both sets: $43.05$ and $42.23$, exceeding Swift-Net-12 by $1.26$ and $2.19$ points, respectively. Further perceptual evaluation and the effect of training data are analyzed in Section~\ref{sec:generalization}. The demo page\footnote{\url{https://cslikai.cn/Falcon/}} provides examples from the four RealSSA-Real scenarios.

\begin{table}[t]
\centering
\small
\caption{WER (\%, $\downarrow$) and target-speaker similarity (cosine similarity $\times 100$, $\uparrow$). Ground truth evaluates the target waveform using the same protocol; ``--'' denotes unavailable isolated target waveforms. Bold and underline mark the best and second-best extraction systems, excluding ground truth.}
\label{tab:wer-speaker-similarity}
\resizebox{0.82\linewidth}{!}{%
\begin{tabular}{lcccc}
\toprule
\multirow{2}{*}{Method} & \multicolumn{2}{c}{RealSSA-Sim} & \multicolumn{2}{c}{RealSSA-Real} \\
\cmidrule(lr){2-3} \cmidrule(lr){4-5}
 & WER (\%) $\downarrow$ & Similarity $\uparrow$ & WER (\%) $\downarrow$ & Similarity $\uparrow$ \\
\midrule
AV-Sepformer~[\citeyear{lin2023av}] & 32.64 & 38.69 & 34.61 & 38.25 \\
AV-TF-GridNet~[\citeyear{pan2023scenario}] & \underline{27.31} & 40.44 & 34.95 & 39.88 \\
RTFSNet-12~[\citeyear{pegg2024rtfsnet}] & 31.91 & 36.51 & 35.56 & 35.78 \\
CTCNet~[\citeyear{li2024audio}] & 34.44 & 37.08 & 38.81 & 35.33 \\
AV-Mossformer2~[\citeyear{zhao2025clearervoice}] & 32.44 & 37.34 & \textbf{32.00} & 36.42 \\
Swift-Net-12~[\citeyear{Sang2025AFA}] & 28.59 & \underline{41.79} & 33.72 & \underline{40.04} \\
Dolphin~[\citeyear{li2026efficient}] & 49.48 & 24.98 & 39.06 & 35.95 \\
\midrule
\textit{Ground truth} & 11.83 & 64.69 & -- & -- \\
\textbf{Falcon (Ours)} & \textbf{25.47} & \textbf{43.05} & \underline{32.74} & \textbf{42.23} \\
\bottomrule
\end{tabular}%
}
  \vspace{-8pt}
\end{table}

\subsection{Computational Efficiency}

We evaluated Falcon's computational efficiency by comparing it with the four strongest causal baselines on RealSSA-Sim. Table~\ref{tab:efficiency} reports parameter counts, MACs, CPU and GPU inference latency, and memory usage under the common measurement protocol described in Appendix~\ref{app:eval_protocol}.

Falcon and Swift-Net-12 both had $0.5$\,M parameters. At this model size, Falcon required $10.6$\,G MACs compared with $45.3$\,G for Swift-Net-12, a reduction of $76.6\%$, while GPU inference latency decreased from $172.1$\,ms to $51.3$\,ms ($70.2\%$). Compared with AV-TF-GridNet, the strongest baseline on RealSSA-Sim, Falcon achieved higher separation quality with $91.9\%$ fewer parameters, $95.0\%$ fewer MACs, and $91.0\%$ lower GPU latency. Compared with Dolphin, Falcon reduced parameter count, MACs, and GPU latency by $91.1\%$, $41.4\%$, and $41.6\%$, respectively. Falcon also had the lowest memory footprint among the evaluated methods, using $125.3$\,MB of CPU RAM and $364.7$\,MB of GPU memory. Together with the separation results in Table~\ref{tab:main-results}, these comparisons show that Falcon combined high extraction quality with low computational and memory costs.

\begin{table}[h]
  \caption{Efficiency comparison among Falcon and the four best-performing causal AVTSE baselines from Table~\ref{tab:main-results}. Lower is better for all metrics.}
  \label{tab:efficiency}
  \centering
  \resizebox{0.88\textwidth}{!}{%
  \begin{tabular}{lcccccc}
    \toprule
    \multirow{2}{*}{Method} & \multirow{2}{*}{Params (M) $\downarrow$} & \multirow{2}{*}{MACs (G) $\downarrow$} & \multicolumn{2}{c}{Latency (ms) $\downarrow$} & \multicolumn{2}{c}{Memory (MB) $\downarrow$} \\
    \cmidrule(lr){4-5} \cmidrule(lr){6-7}
     & & & CPU & GPU & RAM & GPU Inf. \\
    \midrule
    AV-TF-GridNet~[\citeyear{pan2023scenario}]   & 6.2   & 213.8  & 5646.3  & 567.8  & \underline{136.3} & 1015.8 \\
    AV-Mossformer2~[\citeyear{zhao2025clearervoice}]  & 56.7  & 224.2  & 16670.7 & 114.9  & 330.9 & 539.8 \\
    Swift-Net-12~[\citeyear{Sang2025AFA}]        & \textbf{0.5} & 45.3   & 4163.8  & 172.1  & 482.3 & 411.7 \\
    Dolphin~[\citeyear{li2026efficient}]         & \underline{5.6} & \underline{18.1} & \underline{2411.0} & \underline{87.9} & 560.2 & \underline{390.3} \\
    \midrule
    \textbf{Falcon (Ours)} & \textbf{0.5} & \textbf{10.6} & \textbf{1361.2} & \textbf{51.3} & \textbf{125.3} & \textbf{364.7} \\
    \bottomrule
  \end{tabular}%
  }
  \vspace{-8pt}
\end{table}

\Needspace{6\baselineskip}
\subsection{Generalization to Real-World Scenes}
\label{sec:generalization}

\begin{wraptable}[11]{r}{0.55\textwidth}
  \centering
  \footnotesize
  \setlength{\tabcolsep}{4pt}
  \caption{Evaluation on RealSSA-Real when models were trained on VoxCeleb2 versus RealSSA.}
  \label{tab:dataset-ablation}
  \begin{tabular}{lcccc}
    \toprule
    \multirow{2}{*}{Method}
      & \multicolumn{2}{c}{VoxCeleb2}
      & \multicolumn{2}{c}{RealSSA-Sim (Ours)} \\
    \cmidrule(lr){2-3} \cmidrule(lr){4-5}
      & NISQA & DNSMOS & NISQA & DNSMOS \\
    \midrule
    AV-TF-GridNet   & \textbf{2.495} & 2.757 & 2.413 & \textbf{3.030} \\
    AV-Mossformer2  & 1.995 & 2.715 & \textbf{2.256} & \textbf{2.912} \\
    Swift-Net-12    & 2.481 & 3.064 & \textbf{3.484} & \textbf{3.276} \\
    Dolphin         & 1.593 & 2.608 & \textbf{2.590} & \textbf{2.960} \\
    \textbf{Falcon (Ours)} & 2.431 & 2.878 & \textbf{3.615} & \textbf{3.276} \\
    \bottomrule
  \end{tabular}
\end{wraptable}

We compared training on RealSSA-Sim and VoxCeleb2, the largest existing AVTSE corpus, using non-intrusive perceptual quality on the held-out RealSSA-Real.

Table~\ref{tab:dataset-ablation} reports NISQA~\cite{mittag2021nisqa} and DNSMOS~\cite{reddy2022dnsmos} for five causal architectures. RealSSA-Sim training improved DNSMOS for all five and NISQA for four (Falcon: $2.431\!\to\!3.615$ NISQA, $2.878\!\to\!3.276$ DNSMOS). Whereas VoxCeleb2 does not systematically vary near-/far-field interference, environmental and sudden noise, or reverberation, RealSSA-Sim varies these factors jointly. Gains across architectures support systematic acoustic coverage as a basis for real-world generalization.

\begingroup
\setlength{\intextsep}{4pt}
\Needspace{6\baselineskip}
\subsection{Ablation Study}
\label{sec:ablation}

\textbf{Number of downsampling stages.}
The depth of the single-pass separator controls how many time-frequency scales are exploited. Increasing depth from one to three stages improved SI-SNRi by $0.6$\,dB (Table~\ref{tab:downsample-ablation}): the deeper encoder captures broader spectro-temporal context while skip connections progressively recover fine detail. 

\Needspace{9\baselineskip}
\begin{wraptable}{r}{0.42\textwidth}
  \centering
  \footnotesize
  \setlength{\tabcolsep}{4pt}
  \caption{Effect of downsampling depth.}
  \label{tab:downsample-ablation}
  \begin{tabular}{lcc}
    \toprule
    Stages & SI-SNRi $\uparrow$ & SDRi $\uparrow$ \\
    \midrule
    1                  & 7.1 & 7.7 \\
    2                  & \underline{7.4} & \underline{7.9} \\
    \midrule
    \textbf{3 (Ours)}  & \textbf{7.7} & \textbf{8.2} \\
    \bottomrule
  \end{tabular}
\end{wraptable}

Three stages is also the maximum depth permitted by streaming causality, since the visual branch imposes a fixed algorithmic lookahead of $120$\,ms and the audio hop is $8$\,ms, leaving the audio branch at most $15$ frames of lookahead; a fourth stage would exceed this budget and break audio-visual alignment. We used LSTM as the recurrent unit inside each TFRB based on a comparison with RNN, GRU, and SRU reported in Appendix~\ref{app:tfrb_rnn}.

\Needspace{15\baselineskip}
\begin{wraptable}{r}{0.48\textwidth}
  \centering
  \small
  \setlength{\tabcolsep}{4pt}
  \caption{Encoder-space speaker contrastive loss. Gains are consistent across three architectures with different separator designs.}
  \label{tab:loss-ablation}
  \begin{tabular}{llcc}
    \toprule
    Model & Loss & SI-SNRi $\uparrow$ & SDRi $\uparrow$ \\
    \midrule
    Falcon       & Without & 7.3 & 7.8 \\
    Falcon       & +~Ours  & \textbf{7.7} & \textbf{8.2} \\
    \midrule
    SwiftNet-12  & Baseline & 7.3 & 7.8 \\
    SwiftNet-12  & +~Ours   & \underline{7.5} & \underline{8.0} \\
    \midrule
    Dolphin      & Baseline & 6.7 & 7.3 \\
    Dolphin      & +~Ours   & 7.0 & 7.6 \\
    \bottomrule
  \end{tabular}
\end{wraptable}

\textbf{Encoder-space speaker contrastive loss.}
We assessed the proposed loss by removing it from Falcon and by applying it to two recent causal AVTSE models with different separator designs. The contrastive loss improved Falcon by $0.4$\,dB SI-SNRi, SwiftNet-12 by $0.2$\,dB, and Dolphin by $0.3$\,dB (Table~\ref{tab:loss-ablation}). Consistent gains across three architectures suggest that pulling the estimate toward the clean reference and pushing it away from near-field interferers in the encoder feature space provides an architecture-independent regularization signal. The contrastive branch is active only during training, so these gains come at zero inference cost. We further visualized the learned encoder-space speaker embeddings in Appendix~\ref{app:loss-visualization}, showing that the proposed loss yields more speaker-discriminative representations.

\Needspace{12\baselineskip}
\begin{wraptable}{r}{0.48\textwidth}
  \centering
  \small
  \setlength{\tabcolsep}{4pt}
  \caption{Sensitivity to the $\tau$ schedule. Moderate annealing ($0\!\to\!1$) balances early exploration and late hard-negative focus.}
  \label{tab:beta-ablation}
  \begin{tabular}{lcc}
    \toprule
    $\tau$ Strategy & SI-SNRi $\uparrow$ & SDRi $\uparrow$ \\
    \midrule
    Fixed $\tau{=}0$            & \underline{7.5} & \underline{8.1} \\
    Fixed $\tau{=}1$            & 7.4 & 7.9 \\
    Anneal $0\!\to\!2$          & \underline{7.5} & 8.0 \\
    \midrule
    \textbf{Anneal $0\!\to\!1$ (Ours)} & \textbf{7.7} & \textbf{8.2} \\
    \bottomrule
  \end{tabular}
\end{wraptable}

We also examined the annealing schedule of the negative-weighting temperature $\tau$ (Table~\ref{tab:beta-ablation}). Annealing $\tau$ from $0$ to $1$ outperformed both fixed schedules and a more aggressive $0\!\to\!2$ anneal. Uniform initial weighting lets the embedding space stabilize before focus shifts toward hard negatives; fixing $\tau{=}1$ concentrates on a few similar interferers too early, and annealing to $2$ over-concentrates late in training and weakens constraints on moderately similar speakers.

Additional experiments examined output parameterization (Appendix~\ref{app:output_formulation}), speech-overlap ratio (Appendix~\ref{app:overlap_ratio}), and missing visual information (Appendix~\ref{app:visual_missing}).

\endgroup

\section{Conclusions}

RealSSA combines RealSSA-Sim and RealSSA-Real to benchmark AVTSE under variable speaker counts, near-/far-field geometry, reverberation, and sudden noise, with held-out real recordings for evaluating generalization. Falcon performs multi-scale time-frequency separation in a single encoder--decoder pass and uses an encoder-space speaker contrastive loss that suppresses near-field leakage, transfers across backbones, and adds no inference cost. Across RealSSA-Sim, RealSSA-Real, LRS2-2Mix, LRS3-2Mix, and VoxCeleb2-2Mix, Falcon matches or surpasses strong causal baselines. Relative to causal AV-TFGridNet, it reduces parameters by 91.9\%, computation by 20.2$\times$, and GPU inference latency by 11.1$\times$, supporting progress toward low-latency on-device AVTSE.

\textbf{Limitations.} Both the separator and the encoder-space speaker contrastive loss in Falcon operate on a single channel, and we leave the four-element array simulated in Appendix~\ref{app:realssa:pipeline} to future multi-channel extensions. RealSSA-Real covers approximately $84$ minutes over $10$ participants and eight environments, which is sufficient to reveal cross-architecture trends and can be scaled up in follow-up releases. The contrastive loss currently relies on clean per-source waveforms of near-field interferers available in simulation, and adapting it to unlabeled real recordings is a natural direction for future work.

\begin{ack}
This work is supported by the Fundamental and Interdisciplinary Disciplines Breakthrough Plan of the Ministry of Education of China (No.~JYB2025XDXM504), the National Natural Science Foundation of China (No.~62576187), the major science and technology projects in Qinghai Province (2024-GX-A3), and the National Key R\&D Program of China under Grant Nos.~2024YFB2808800 and 2024YFB2808803.
\end{ack}

\bibliographystyle{unsrtnat}
\bibliography{references}


\appendix

\section{Detailed Dataset Construction of RealSSA}
\label{app:realssa}

This appendix provides the construction details of RealSSA that complement Section~\ref{sec:srav}. We describe the source corpora and the rationale behind the four scene types (Appendix~\ref{app:realssa:source}), the spatial simulation and mixing pipeline (Appendix~\ref{app:realssa:pipeline}), the resulting distribution statistics (Appendix~\ref{app:realssa:stats}), and RealSSA-Real (Appendix~\ref{app:realssa-real}). All design choices are made to make the benchmark reproducible from public corpora and to expose AVTSE models to acoustic conditions that fixed two-speaker datasets do not contain.

\subsection{Source Data and Scene Design}
\label{app:realssa:source}

\paragraph{Source corpora.}
RealSSA-Sim is built from four publicly available corpora that play complementary roles in the mixture (Table~\ref{tab:srav_sources}). LRS3~\cite{afouras2018lrs3} provides target audio together with the temporally aligned mouth video that AVTSE models rely on, and is the only corpus among the four that supplies paired audio-visual data at scale. LibriSpeech~\cite{panayotov2015librispeech} contributes near-field and far-field interfering speakers; its large speaker inventory and clean read speech allow us to add interferers without re-introducing the noise already present in the target corpus. WHAM!~\cite{wichern2019wham} supplies continuous environmental backgrounds recorded in real urban scenes, which we use to populate the always-on noise floor of each mixture. Hive~\cite{li2026semantically} is a recently released sound-event corpus whose semantic label hierarchy lets us draw scene-appropriate environmental and sudden-noise events; this hierarchy is the property that distinguishes RealSSA-Sim from prior simulated AVTSE sets that mix WHAM! noise without semantic control. LRS3, LibriSpeech, and WHAM! are natively at $16\,\mathrm{kHz}$, while Hive is provided at $44.1\,\mathrm{kHz}$ and is resampled to $16\,\mathrm{kHz}$ before mixing.

\begin{table}[h]
\centering
\caption{Source corpora used to construct RealSSA-Sim. 
``Role'' summarizes the function of each corpus in the mixture. 
``Hours'' denotes the approximate total speech duration. 
``Speakers'' indicates the number of distinct speakers (marked ``--'' if not applicable). 
``Native SR'' is the original sampling rate, and ``Used SR'' is the resampled rate used in RealSSA-Sim.}
\label{tab:srav_sources}
\resizebox{\linewidth}{!}{
\begin{tabular}{llcccc}
\toprule
Corpus & Role & Hours & Speakers & Native SR & Used SR \\
\midrule
LRS3~\cite{afouras2018lrs3}            & Target (audio--visual)               & $\sim$438       & $\sim$9{,}545  & $16\,\mathrm{kHz}$    & $16\,\mathrm{kHz}$ \\
LibriSpeech~\cite{panayotov2015librispeech} & Near-/far-field interferers     & $\sim$1000      & $\sim$2{,}484  & $16\,\mathrm{kHz}$    & $16\,\mathrm{kHz}$ \\
WHAM!~\cite{wichern2019wham}           & Continuous environmental noise       & $\sim$80        & --             & $16\,\mathrm{kHz}$    & $16\,\mathrm{kHz}$ \\
Hive~\cite{li2026semantically}         & Semantic environmental and sudden events & $\sim$544 & --             & $44.1\,\mathrm{kHz}$ & $16\,\mathrm{kHz}$ \\
\bottomrule
\end{tabular}}
\end{table}

\paragraph{Scene design rationale.}
The four scene types in Table~\ref{tab:srav_config} are not arbitrary. They are chosen to cover an acoustic gradient that separators must traverse in deployment, with three controlled axes: room volume, reverberation time, and the presence of far-field interferers. Small meeting rooms have the shortest RT60 (0.3--0.6\,s), match the scale of two-person to small-group offices, and contain only near-field interferers because no point in the room is beyond the 3\,m near-/far-field threshold; this makes them the ``clean'' end of the gradient. Medium indoor spaces (e.g.,\ kitchens, living rooms) introduce moderate reverberation (0.5--0.9\,s) together with the first far-field talkers, since the room becomes large enough for some speakers to fall outside the critical-distance neighborhood of the microphone~\cite{kuttruff2016room}. Large public spaces such as cafeterias and lecture halls extend RT60 to 1.0--2.0\,s, which is the regime in which reverberant energy dominates the direct path and in which far-field speech becomes the harder interferer to suppress. Outdoor environments invert the indoor pattern: very low RT60 (0.05--0.1\,s) but high environmental SIR difficulty, so that the stress test moves from reverberation to wide-band ambient noise. Together the four scenes span the regime in which residential and on-the-go AVTSE systems are deployed, and the RT60 ranges fall within the values reported for occupied rooms of comparable size in standard architectural-acoustics references~\cite{kuttruff2016room}.

\begin{figure}[h]
\centering
\begin{subfigure}[b]{0.48\linewidth}
  \centering
  \includegraphics[width=\linewidth]{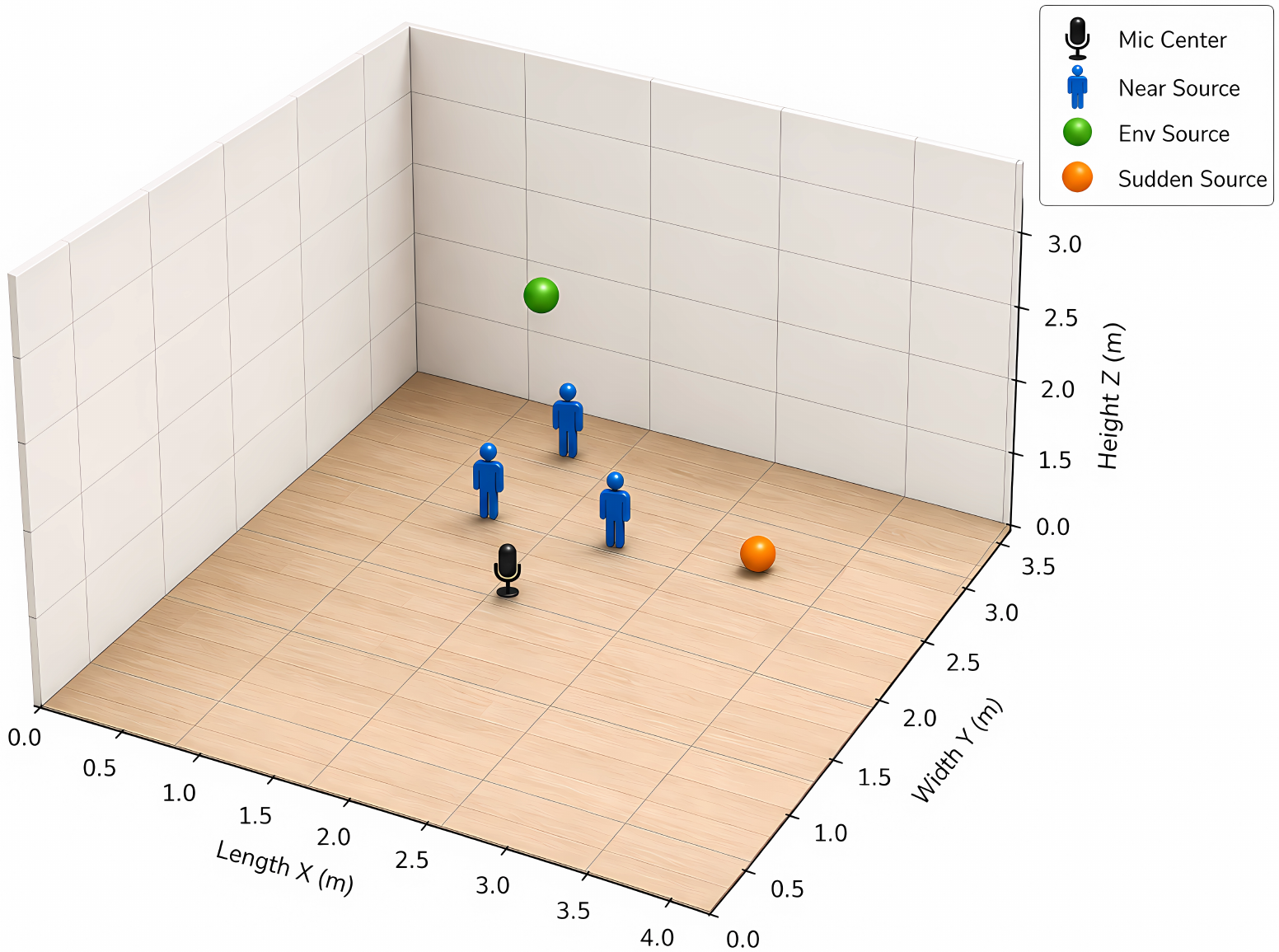}
  \caption{Small meeting}
\end{subfigure}
\hfill
\begin{subfigure}[b]{0.48\linewidth}
  \centering
  \includegraphics[width=\linewidth]{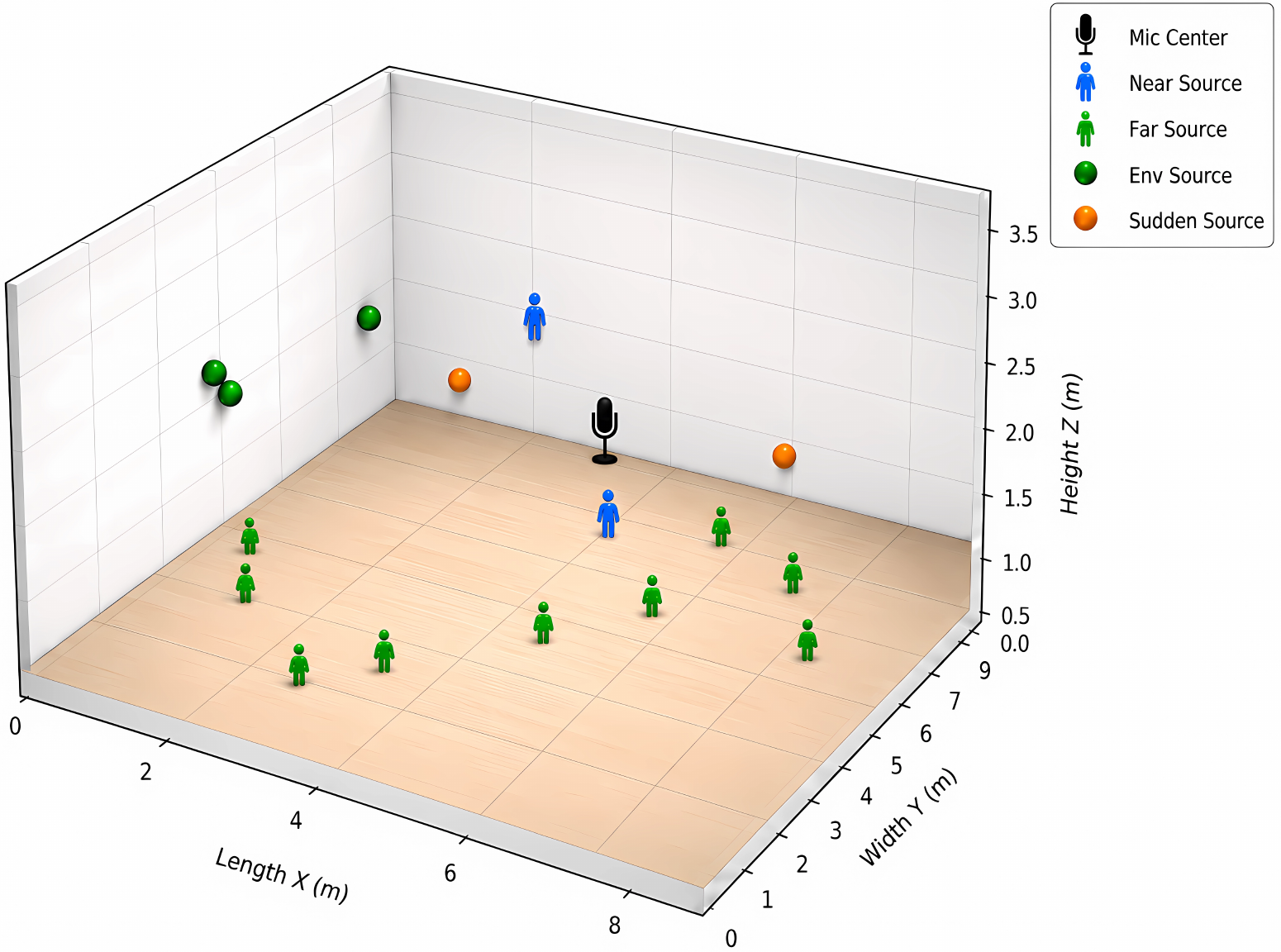}
  \caption{Medium indoor}
\end{subfigure}

\begin{subfigure}[b]{0.48\linewidth}
  \centering
  \includegraphics[width=\linewidth]{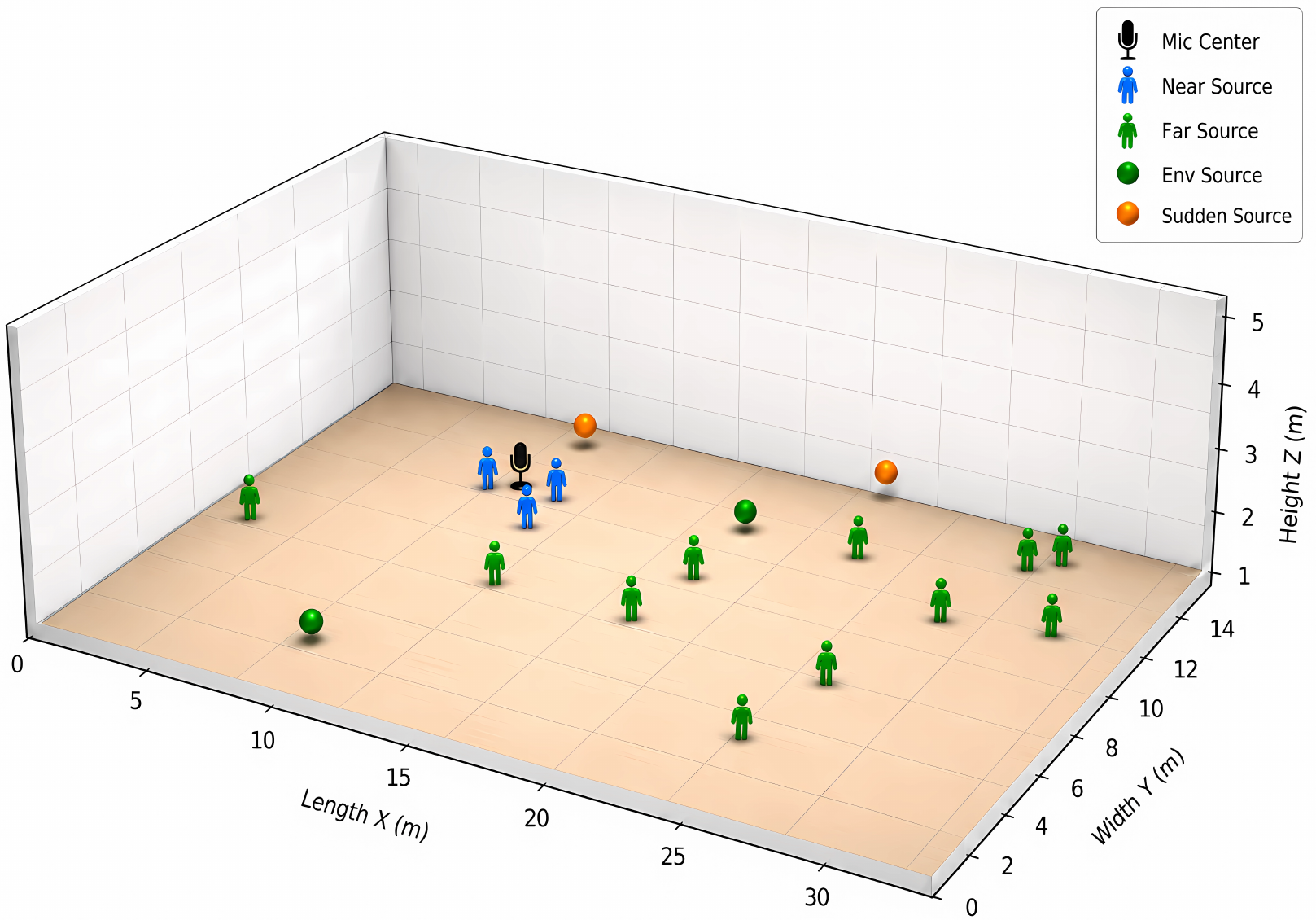}
  \caption{Large public}
\end{subfigure}
\hfill
\begin{subfigure}[b]{0.48\linewidth}
  \centering
  \includegraphics[width=\linewidth]{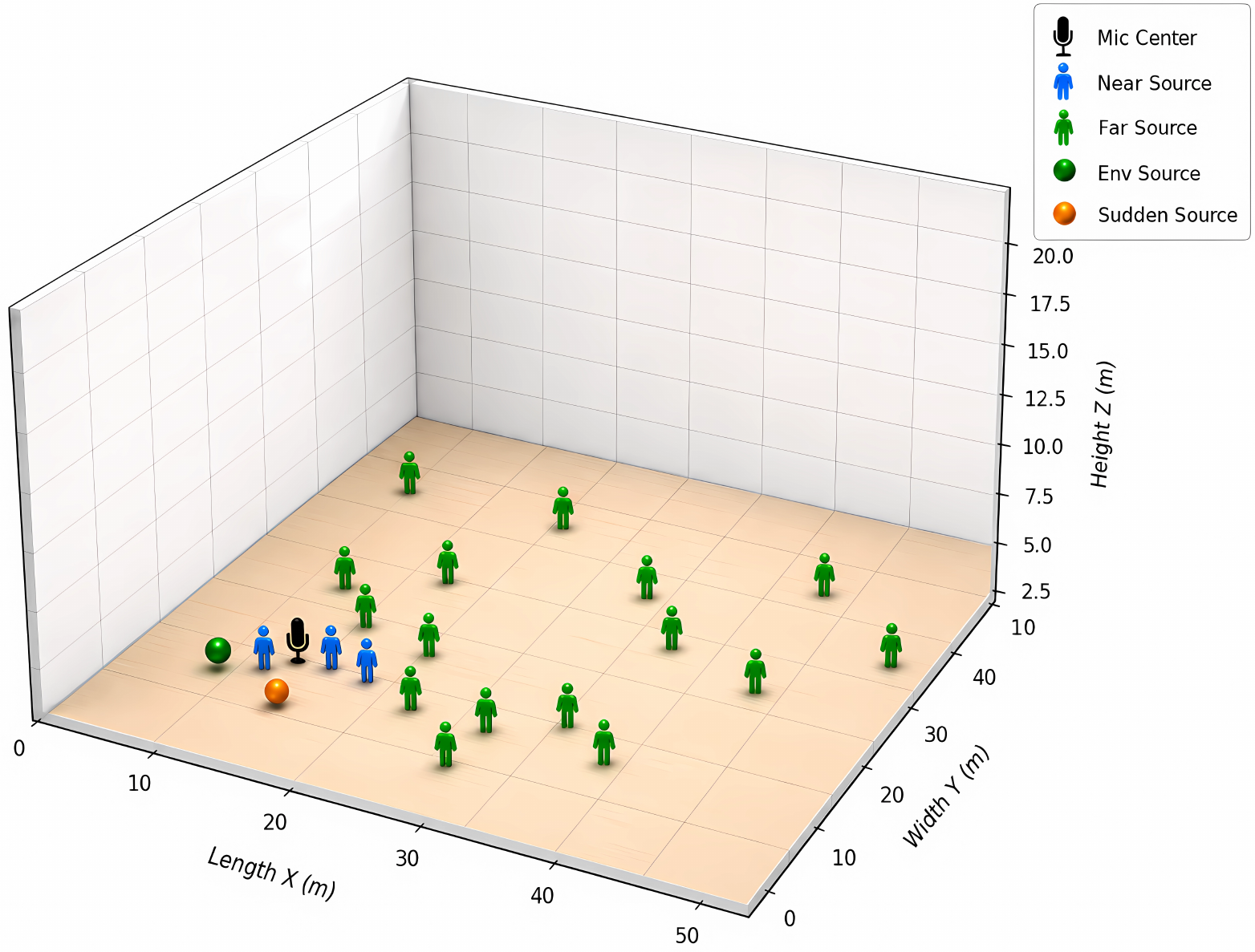}
  \caption{Outdoor}
\end{subfigure}
\caption{Spatial layouts of the four RealSSA-Sim acoustic scene types. The blue human icons indicate the positions of near-field speakers, while the green human icons indicate the positions of far-field speakers. The green spheres denote environmental sound sources, the orange spheres denote sudden noise sources, and the black microphone denotes the microphone position. Since different sound sources may have different spatial locations, the visualization icons can appear at different heights.}
\label{fig:scene_layouts}
\end{figure}

\paragraph{Near-field overlap design.}
The overlap-ratio range of 10\%--40\% used in small-meeting samples is grounded in the speaker-overlap statistics measured in conversational corpora, where the fraction of dyadic-conversation time with simultaneous speech typically falls in this band. We therefore sample partial-overlap intervals only in the small-meeting scene, where short conversational turn-taking is plausible, and apply full-length alignment in larger scenes where the dominant interference pattern is co-occurring background talkers rather than turn-taking. The extended scene-parameter table is given in Table~\ref{tab:srav_config_full}; it reproduces the four columns of the in-text Table~\ref{tab:srav_config} and additionally specifies the per-scene Near SIR ranges and the applied far-field SIR range of $20$--$35\,\mathrm{dB}$ for scenes with far-field interferers.

\begin{table}[h]
\centering
\caption{Extended acoustic-scene configuration for RealSSA-Sim. Near speakers and Far speakers denote interfering speakers and exclude the target. Near SIR specifies the scene-dependent range for near-field interferers. Applied Far SIR is $20$--$35\,\mathrm{dB}$ in all scenes with far-field interferers; small-meeting scenes contain none. The mixing procedure is detailed in Appendix~\ref{app:realssa:pipeline}.}
\label{tab:srav_config_full}
\resizebox{\linewidth}{!}{
\begin{tabular}{lcccccccc}
\toprule
Scene & Space size & RT60 & Near speakers & Far speakers & Near SIR & Env. noise SIR & Applied Far SIR & Sudden noise \\
\midrule
Small meeting
& \(4\text{-}6 \times 3\text{-}5 \times 2.5\text{-}3\,\mathrm{m}\)
& 0.3--0.6\,s
& 1--5, 0.5--1.5\,m
& none
& 2--5\,dB
& 5--15\,dB
& --
& 30\%, 1 event \\
Medium indoor
& \(8\text{-}15 \times 8\text{-}12 \times 3\text{-}4\,\mathrm{m}\)
& 0.5--0.9\,s
& 1--3, 1.0--2.5\,m
& 10--15, 3--15\,m
& 2--8\,dB
& 5--10\,dB
& 20--35\,dB
& 40\%, 1--2 events \\
Large public
& \(20\text{-}40 \times 15\text{-}30 \times 4\text{-}8\,\mathrm{m}\)
& 1.0--2.0\,s
& 1--4, 1.0--3.0\,m
& 10--15, 5--20\,m
& 2--8\,dB
& 2--8\,dB
& 20--35\,dB
& 50\%, 1--3 events \\
Outdoor
& \(50 \times 50 \times 20\,\mathrm{m}\)
& 0.05--0.1\,s
& 1--4, 1.0--4.0\,m
& 10--15, 5--25\,m
& 2--15\,dB
& $-2$ to $5$\,dB
& 20--35\,dB
& 40\%, 1--3 events \\
\bottomrule
\end{tabular}}
\end{table}

\paragraph{Semantic noise selection from Hive.}
Hive provides a label hierarchy over sound events; we use it to ensure that the noise pool of each scene is semantically consistent with the scene type rather than uniformly drawn. Table~\ref{tab:srav_hive_labels} lists the environmental and sudden-noise labels assigned to each scene. The mapping follows three rules: continuous mechanical noise (air conditioning, refrigerator, microwave) is restricted to indoor scenes whose physical setting plausibly contains such sources; pedestrian and traffic events (footsteps, sirens, vehicle pass-bys) are restricted to large-public and outdoor scenes; and a small pool of generic transient events (key jangling, coin dropping, slam, bang) is shared across all scenes since they occur ubiquitously regardless of room type. The label-to-scene assignment is fixed in the released metadata and is the only step in the pipeline that draws on human knowledge of the sound-event taxonomy.

\begin{table}[h]
\centering
\caption{Hive label assignment per RealSSA-Sim scene. ``All scenes (shared)'' lists generic transient events that are added to the sudden-noise pool of every scene.}
\label{tab:srav_hive_labels}
\renewcommand{\arraystretch}{1.15}
\begin{tabularx}{\linewidth}{@{}l X X@{}}
\toprule
Scene & Environmental noise labels & Sudden-noise labels \\
\midrule
Small meeting     & air conditioning, fan, clock, typing, printer, keyboard, tick, hiss
                  & writing, clicking, crumple, rustle, knock, telephone, door, drawer, squeak, paper \\
Medium indoor     & microwave, blender, dishes, pots/pans, refrigerator
                  & cutlery, glass, cupboard, frying, cash register, gurgling, steam, kettle, sneeze, cough, laughter \\
Large public      & footsteps, shuffle, speech, babble, crowd
                  & sliding door, siren, child speech, crying, roll, alarm, buzzer, announcement \\
Outdoor           & wind, rain, traffic, stream, city ambience
                  & car, bus, truck, motorcycle, car passing, skidding, reversing beep, horn, police/ambulance/fire siren, dog, bird, cricket, jackhammer, hammer, power tool \\
All scenes (shared) & --
                  & key jangling, coin dropping, slam, bang, thump, crash, squeal \\
\bottomrule
\end{tabularx}
\end{table}

\subsection{Generation Pipeline}
\label{app:realssa:pipeline}

\paragraph{Spatial simulation.}
Room impulse responses (RIRs) are computed with gpuRIR~\cite{diaz2021gpurir} using the image-source method. For every sample, we draw a room geometry and an RT60 target from the ranges in Table~\ref{tab:srav_config_full}, then convert the RT60 target into image-source attenuation parameters using Sabine's estimator at the gpuRIR cut-offs of $15\,\mathrm{dB}$ and $60\,\mathrm{dB}$, which controls when image sources stop contributing direct and reverberant energy. The microphone is modelled as a four-element linear array with $2.8\,\mathrm{cm}$ spacing whose array centre is placed at a height of $1.2$--$1.8\,\mathrm{m}$ above the floor; this geometry mirrors the form factor of speakerphones and laptop microphones used in conferencing devices. Although the array is generated to enable future multi-channel extensions, all experiments in the main paper use the first-channel signal only, so RealSSA in its current form is a single-channel benchmark and the released waveforms are the channel-1 streams of the simulated array. Speaker heights are sampled from $1.0$--$1.9\,\mathrm{m}$ to span the range of seated and standing posture.

\paragraph{Outdoor simulation.}
Outdoor scenes are approximated by a $50\times50\times20\,\mathrm{m}$ enclosure with wall absorption coefficient $0.99$, which yields RT60 in $0.05$--$0.1\,\mathrm{s}$ and produces an essentially anechoic propagation profile in which the direct path dominates. We adopt this approximation because gpuRIR's image-source model is constructed for closed rooms and does not natively model open-field propagation. The simplification preserves the inverse-distance amplitude law and the geometric arrival times of all sources, which are the dominant cues that an AVTSE model uses, and removes only late reverberation, which is in any case negligible in true outdoor scenes. We make this assumption explicit so that downstream users of RealSSA can interpret outdoor results accordingly.

\paragraph{Mixing pipeline.}
Algorithm~\ref{alg:srav_pipeline} formalizes the per-sample mixing procedure. The target speech is first convolved with the target RIR and loudness-normalized to $-23\,\mathrm{LUFS}$ following the EBU R128 broadcast loudness recommendation~\cite{ebu2020r128}, which is the standard reference loudness for speech-centric content and removes the residual loudness variability of LRS3. Near-field interferers are added at the per-scene Near SIR ranges in Table~\ref{tab:srav_config_full}. For medium-indoor, large-public, and outdoor scenes, far-field interferers are scaled to an applied SIR of $20$--$35\,\mathrm{dB}$, as listed in the Applied Far SIR column. Small-meeting scenes contain no far-field interferers. Continuous environmental noise is the sum of one WHAM! stream and one Hive ambient stream; when both are present we attenuate each by $3\,\mathrm{dB}$ before summation so that the total noise SIR matches the per-scene environmental-noise SIR specification rather than doubling. Sudden-noise events are inserted at uniformly-sampled non-overlapping positions according to the per-scene event count, with each event scaled to the same SIR distribution as the environmental noise. The final mixture is peak-limited at $-1\,\mathrm{dBFS}$ to prevent digital clipping, and per-source gains are clipped to the range $[-40, +40]\,\mathrm{dB}$ to suppress the rare numerical instability that can arise when an exceptionally quiet source is paired with a high target SIR.

\begin{algorithm}[h]
\caption{RealSSA-Sim per-sample mixture generation.}
\label{alg:srav_pipeline}
\begin{algorithmic}[1]
\Require Target utterance $\mathbf{s}$ with paired mouth video $\mathbf{m}$ from LRS3
\Require Scene type $c \in \{\text{small\_meeting}, \text{medium\_indoor}, \text{large\_public}, \text{outdoor}\}$
\State Sample room geometry, RT60, and source/microphone positions from the ranges of scene $c$
\State Compute RIRs $\{\mathbf{h}_{\text{s}}, \mathbf{h}_1, \dots, \mathbf{h}_K\}$ with gpuRIR and Sabine attenuation
\State Convolve target with $\mathbf{h}_{\text{s}}$ and normalize the result to $-23\,\mathrm{LUFS}$
\State Add near-field LibriSpeech interferers at Near SIR; partial overlap of $10\%$--$40\%$ for small\_meeting, full-length alignment otherwise
\State For scenes with far-field speakers, add LibriSpeech interferers at an applied SIR of $20$--$35\,\mathrm{dB}$
\State Add WHAM! continuous noise and one Hive ambient stream at the scene-specific environmental SIR (each attenuated by $3\,\mathrm{dB}$ when both are present)
\State Insert per-scene Hive sudden-noise events at random non-overlapping positions
\State Apply per-source gain clipping to $[-40, +40]\,\mathrm{dB}$
\State Peak-limit the mixture at $-1\,\mathrm{dBFS}$
\State \Return mixture waveform, clean target, mouth video, and source-/scene-level metadata
\end{algorithmic}
\end{algorithm}

\paragraph{Reproducibility.}
All randomness is controlled by a fixed master seed of $42$, with disjoint seed offsets for the training, validation, and test partitions so that re-running the pipeline regenerates the released splits bit-for-bit. The released metadata records the sampled room geometry, RT60, per-source positions, RIR hashes, SIR values, and Hive label identifiers for each sample. The pipeline depends on \texttt{gpuRIR}, \texttt{pyloudnorm}, \texttt{torchaudio}, and \texttt{soundfile}; on a single NVIDIA RTX\,3090 the full $100{,}826$-sample training set takes approximately 48h, GPU-hours to generate, which is dominated by the gpuRIR convolution stage. We will release the generation scripts and the metadata together with the dataset.

\subsection{Dataset Statistics}
\label{app:realssa:stats}

\begin{table}[h]
\centering
\caption{Per-split summary statistics of RealSSA-Sim. Total number of speakers includes the target and all near-/far-field interferers. Number of speakers, near-/far-field SIR, and RT60 are sampled per mixture from the ranges of Table~\ref{tab:srav_config_full}; the values reported here are realized averages over the released metadata.}
\label{tab:srav_split_stats}
\resizebox{\linewidth}{!}{
\begin{tabular}{lcccccc}
\toprule
Split & Samples & Total hours & Avg.\ duration & Avg.\ speakers & Avg.\ Near SIR & Avg.\ Applied Far SIR \\
\midrule
Train & $100{,}826$ & $129.2$ & $4.61$\,s & $13.21$ & $8.51$\,dB & $27.49$\,dB \\
Val   & $9{,}133$   & $11.7$  & $4.61$\,s & $13.29$ & $8.51$\,dB & $27.47$\,dB \\
Test  & $9{,}169$   & $11.7$  & $4.61$\,s & $13.23$ & $8.49$\,dB & $27.50$\,dB \\
\bottomrule
\end{tabular}}
\end{table}

\begin{figure}[h]
\centering
\includegraphics[width=0.95\linewidth]{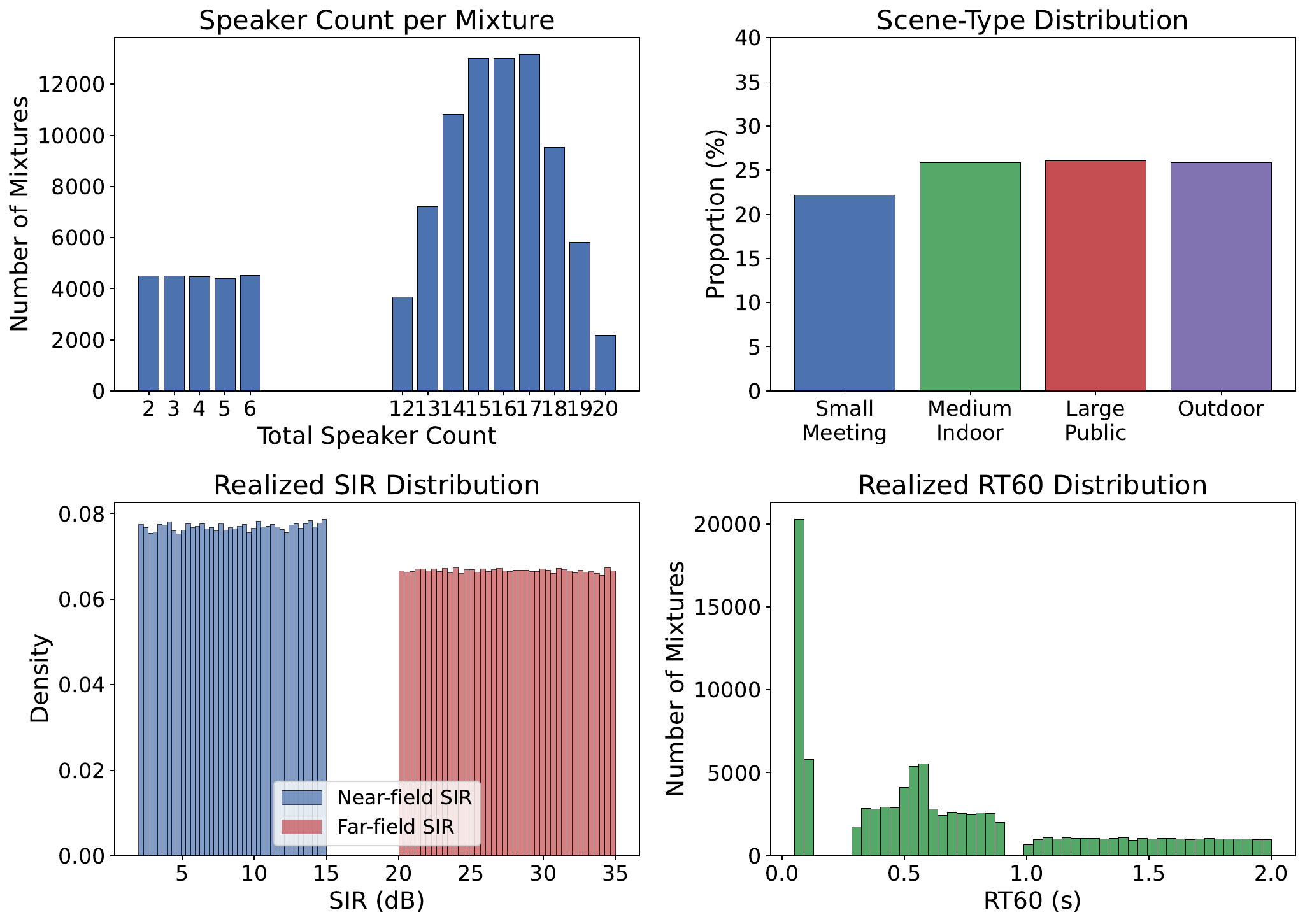}
\caption{RealSSA-Sim distribution statistics. (a) Total number of speakers per mixture, including the target and all interferers. (b) Scene-type proportion. (c) Realized near-field and far-field SIR distributions. (d) Realized RT60 distribution. Histograms are computed from the released training-split metadata.}
\label{fig:srav_dist}
\end{figure}

Table~\ref{tab:srav_split_stats} summarizes the realized statistics of the three splits. The four scene types are sampled in approximately balanced proportions, with each accounting for nearly one quarter of every split, which is reflected in the scene-distribution panel of Figure~\ref{fig:srav_dist}. Figure~\ref{fig:srav_dist} additionally presents aggregate per-mixture distributions of the total number of speakers, the realized near-field and far-field SIR, and the RT60 of the simulated rooms. The RT60 histogram exhibits the four-mode pattern induced by the four scene RT60 bands, with the outdoor mode tightly concentrated near $0.05\,\mathrm{s}$ and the large-public mode spread between $1.0$ and $2.0\,\mathrm{s}$.

\subsection{RealSSA-Real: Real-World Evaluation Set}
\label{app:realssa-real}

\paragraph{Recording protocol.}
RealSSA-Real complements the simulated suite and is collected separately, so that no participant or location is reused between the two. Each session is captured by a single Apple iPhone~17~Pro, whose rear camera and built-in microphones acquire video and audio in one device pass at $60$\,fps and $48\,\mathrm{kHz}$, respectively. The audio is then downsampled to $16\,\mathrm{kHz}$ and the video to $25$\,fps to match the AVTSE pipeline used during training. The iPhone is placed on a table at conversational height in indoor scenes and held at chest level in outdoor scenes, which keeps the device-to-source geometry within the range used in the simulated suite of Appendix~\ref{app:realssa:pipeline}. We deliberately do not denoise, dereverberate, or volume-normalize the recordings, so that the evaluation reflects the raw on-device signal that a deployed AVTSE system would receive.

\begin{figure}[h]
\centering
\includegraphics[width=\linewidth]{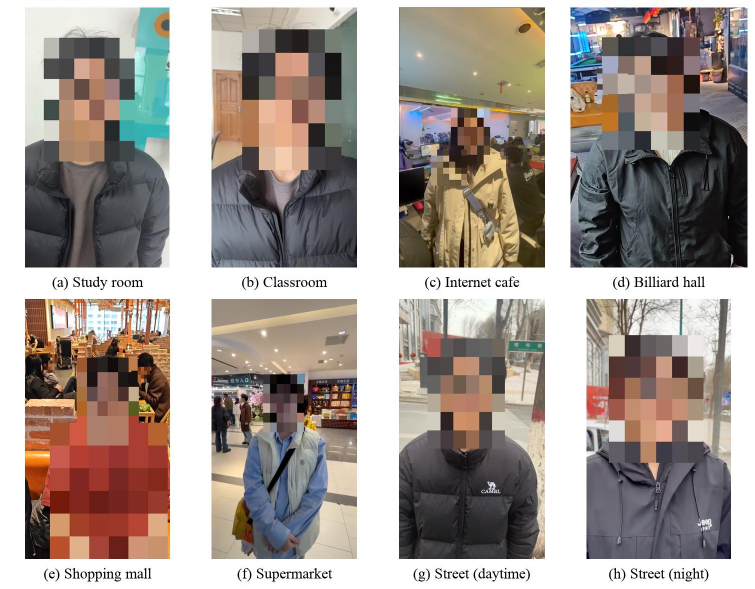}
\caption{Eight environments covered by RealSSA-Real. All recordings were captured with a single Apple iPhone~17~Pro; detected faces have been pixelated for privacy before release.}
\label{fig:realworld_scenes}
\end{figure}

\paragraph{Scenes and contents.}
The set contains $54$ sessions totalling $5{,}035$\,s (approximately $84$ minutes) of overlapping speech across eight environments that span quiet indoor rooms, crowded public venues, and open streets (Figure~\ref{fig:realworld_scenes}): a study room, a classroom, an internet cafe, a billiard hall, a shopping mall, a supermarket, and two street scenes captured in daytime and at night. Per-environment session counts and durations are reported in Table~\ref{tab:realworld_stats}. Recordings involve $10$ participants ($5$ male and $5$ female), with $2$--$5$ participants present in each session. Every participant provided written consent for the recording and for its subsequent academic release, and all participant faces are pixelated prior to release. Each environment carries its own non-stationary noise signature: mall and supermarket sessions combine crowd chatter with venue ambient noise, including noise from nearby diners in the mall and from passing shoppers in the supermarket; billiard-hall sessions add intermittent ball-collision impacts together with background music playing in the venue; internet-cafe sessions contain venue ambient noise layered with crowd chatter; the two street scenes contain vehicular traffic, pedestrian chatter, and occasional music from nearby storefronts; classroom sessions add desk-tapping and footstep events; study-room sessions add coughing and door-opening events. The classroom and study-room sessions each contain $2$ near-field talkers, whereas the remaining public and outdoor environments contain $1$ near-field talker. Sessions are organized so that the number of simultaneous speakers, the speaker positions, and the ambient-noise sources jointly match the scene categories of Table~\ref{tab:srav_config_full}; this matching is used only to organize the test set and is not exposed to the model at inference time.

\begin{table}[h]
\centering
\caption{Per-environment composition of RealSSA-Real. Total durations are reported as ``minutes:seconds''. The environments are grouped by the simulated scene category they correspond to in Table~\ref{tab:srav_config_full}.}
\label{tab:realworld_stats}
\begin{tabular}{llcc}
\toprule
Simulated scene & Real-world environment & Sessions & Total duration (m:s) \\
\midrule
\multirow{2}{*}{Small meeting}  & Study room    & $12$ & $10{:}36$ \\
                                 & Classroom     & $8$  & $10{:}16$ \\
\midrule
\multirow{2}{*}{Medium indoor}  & Internet cafe & $5$  & $10{:}22$ \\
                                 & Billiard hall & $5$  & $10{:}39$ \\
\midrule
\multirow{2}{*}{Large public}   & Shopping mall & $5$  & $9{:}48$  \\
                                 & Supermarket   & $5$  & $10{:}46$ \\
\midrule
\multirow{2}{*}{Outdoor}        & Street (daytime) & $7$ & $9{:}57$  \\
                                 & Street (night)   & $7$ & $11{:}31$ \\
\midrule
Total                            & --            & $54$ & $83{:}55$ \\
\bottomrule
\end{tabular}
\end{table}

\paragraph{References and evaluation metrics.}
We evaluated RealSSA-Real along three complementary dimensions: perceptual quality using DNSMOS~\cite{reddy2022dnsmos} and NISQA~\cite{mittag2021nisqa}, speech recognition using WER with Mega-ASR~\cite{xie2026mega}, and target-speaker similarity using WeSpeaker~\cite{wang2023wespeaker}. RealSSA-Real has no isolated target waveforms. WER and speaker similarity use target transcripts and independent enrollment utterances, respectively, with the protocols detailed in Appendix~\ref{app:eval_protocol}. RealSSA-Real was held out for evaluation and was never used for training or validation.

\paragraph{Ethics statement.}
All recordings were collected under written informed consent. Participants were informed of the academic purpose of the data, were free to withdraw their contribution at any time, and were offered the option to redact specific segments before release. Participants received appropriate compensation for their time, at a rate at or above the local minimum hourly wage. The released package contains only the segments for which release consent was confirmed, and all personally identifiable visual information is pixelated prior to release; no speaker names, contact information, or other identifying metadata are included.

\section{Dataset Setup Details}
\label{app:dataset_setup}

This appendix complements Section~5.1 of the main text and groups the preprocessing details that are shared across the five evaluation sets.

\paragraph{Public AVTSE benchmarks.}
We used LRS2~\cite{afouras2018deep}, LRS3~\cite{afouras2018lrs3}, and VoxCeleb2~\cite{chung2018voxceleb2} under their official train, validation, and test partitions. To form two-speaker mixtures, we followed the protocol of prior causal AVTSE work~\cite{li2024audio,Sang2025AFA}. For every utterance we extracted a fixed-length 2-second clip; clips shorter than 2 seconds were zero-padded and clips longer than 2 seconds were randomly cropped during training. Audio was resampled to $16\,\mathrm{kHz}$, and the paired video was decoded to $25$\,fps so that each clip yielded exactly $T_{\text{v}}=50$ video frames. We cropped a $96\times96$ mouth region of interest around the lip centroid and converted it to grayscale, matching the visual front-end used by~\cite{li2024audio,Sang2025AFA}.

\paragraph{RealSSA-Sim.}
The RealSSA-Sim simulated suite was constructed as described in Section~\ref{sec:srav} and Appendix~\ref{app:realssa}. The training, validation, and test splits contain $100{,}826$, $9{,}133$, and $9{,}169$ samples respectively (Appendix~\ref{app:realssa:stats}). Audio sampling rate, video frame rate, and mouth ROI followed the same configuration as the public benchmarks, so that a single Falcon checkpoint can be evaluated across all five sets without front-end re-tuning.

\paragraph{RealSSA-Real.}
Recording protocol, scene composition, and annotation procedure for RealSSA-Real are documented in Appendix~\ref{app:realssa-real}. The set was held out for evaluation only and was never used for training or validation. The same $96\times96$ mouth ROI extraction and $16\,\mathrm{kHz}$ audio resampling were applied so that the input format matches that of the simulated splits.

\section{Falcon Architecture Widths}
\label{app:arch_widths}

This appendix complements Section~5.2 of the main text and maps every channel-width symbol introduced in Sections~\ref{subsec:overall_pipeline} and~\ref{sec:separator} to its concrete value (Table~\ref{tab:falcon_widths}).

\begin{table}[h]
\centering
\caption{Channel-width settings of Falcon. Symbols follow the definitions in Sections~\ref{subsec:overall_pipeline} and~\ref{sec:separator}.}
\label{tab:falcon_widths}
\begin{tabular}{llc}
\toprule
Symbol & Module & Value \\
\midrule
$C_{\text{e}}$ & STFT encoder output channels & $112$ \\
$C$ & Audio bottleneck output channels & $112$ \\
$C_{\text{h}}$ & Separator hidden channels & $64$ \\
$C_{\text{v}}$ & Visual hidden channels & $32$ \\
$\dim(\mathbf{q})$ & Speaker embedding dimension & $64$ \\
$h_{\mathcal{F}}$ & TFRB frequency-LSTM hidden size & $24$ \\
$h_{\mathcal{T}}$ & TFRB time-LSTM hidden size & $72$ \\
$n_{\text{head}}$ & TFRB attention heads & $4$ \\
$L$ & Separator levels & $4$ \\
\bottomrule
\end{tabular}
\end{table}

\section{Falcon Training Protocol}
\label{app:training_protocol}

Table~\ref{tab:falcon_training} lists the full training configuration. We used the same FRCNN visual backbone as CTCNet~\cite{li2024audio}, initialized from a publicly released lipreading-pretrained checkpoint and frozen throughout AVTSE training. The contrastive temperature $\tau$ was linearly annealed from $0$ to $1$ over the first $1000$ training epochs and held at $1$ thereafter, in line with the schedule of Section~\ref{sec:training} and the analysis of Table~\ref{tab:beta-ablation}. The reconstruction loss $\mathcal{L}_{\text{recon}}$ uses the negative SI-SNR formulation defined in Section~\ref{sec:training}, and the contrastive weight was fixed at $\lambda_{\text{spk}}=1.0$ as reported in Section~\ref{sec:training}.

\begin{table}[h]
\centering
\caption{Training protocol of Falcon.}
\label{tab:falcon_training}
\begin{tabular}{ll}
\toprule
Item & Value \\
\midrule
Optimizer & AdamW, learning rate $1\times10^{-3}$, weight decay $0.01$ \\
LR scheduler & ReduceLROnPlateau, factor $0.5$, patience $15$ \\
Early stopping & Patience $30$ on validation loss \\
Max epochs & $1000$ \\
Batch size per GPU & $4$ \\
Hardware & $8\times$ NVIDIA RTX 3090 \\
Audio segment length & $2\,\mathrm{s}$ at $16\,\mathrm{kHz}$ \\
Video frame rate & $25$\,fps, $T_{\text{v}}=50$ frames \\
Mouth ROI & $96\times96$ grayscale \\
STFT & Window $256$, hop $128$ ($8\,\mathrm{ms}$ at $16\,\mathrm{kHz}$), Hann \\
Encoder input channels & Real, imaginary, magnitude (3 channels) \\
Visual backbone & FRCNN (as in CTCNet), frozen, lipreading-pretrained \\
Reconstruction loss & $\mathcal{L}_{\text{recon}}=-\mathrm{SI\text{-}SNR}(\hat{\mathbf{s}},\mathbf{s})$ \\
Contrastive loss weight $\lambda_{\text{spk}}$ & $1.0$ \\
Contrastive margin $m$ & $0.5$ \\
$\tau$ schedule & Linear $0\to1$ over first $1000$ epochs, held at $1$ thereafter \\
Random seed & $42$ \\
\bottomrule
\end{tabular}
\end{table}

\section{Evaluation Metrics and Inference Latency Protocol}
\label{app:eval_protocol}

\paragraph{Target speech extraction quality.}
On the simulated benchmarks (LRS2-2Mix, LRS3-2Mix, VoxCeleb2-2Mix, and RealSSA-Sim), clean references are available and we reported scale-invariant signal-to-noise ratio improvement (SI-SNRi) and signal-to-distortion ratio improvement (SDRi)~\cite{vincent2006performance}, both in dB, relative to the unprocessed mixture.

\paragraph{Perceptual quality on real recordings.}
We assessed the perceptual quality of extracted speech on RealSSA-Real with DNSMOS~\cite{reddy2022dnsmos} and NISQA~\cite{mittag2021nisqa}. These non-intrusive predictors estimate perceived speech quality directly from the extracted waveform, complementing the speech recognizability and target-speaker consistency measured by WER and speaker similarity, respectively.

\paragraph{Speech recognition.}
We used Mega-ASR~\cite{xie2026mega} to transcribe the extracted speech and compute word error rate (WER) against the target transcript. WER is the number of substituted, deleted, and inserted words divided by the number of reference words, reported as a percentage. Evaluation covered all $9{,}169$ RealSSA-Sim test samples and all $54$ RealSSA-Real recordings.

\paragraph{Target-speaker similarity.}
We extracted speaker embeddings using the ResNet34-LM checkpoint released by WeSpeaker~\cite{wang2023wespeaker}\footnote{Official checkpoint: \url{https://huggingface.co/Wespeaker/wespeaker-voxceleb-resnet34-LM}.}. This model was trained on VoxCeleb2 and further fine-tuned with a large-margin objective. For each test sample, we computed the cosine similarity between the extracted speech and an independent enrollment utterance from the same target speaker. The enrollment utterance differed from the test utterance and was used only for evaluation. We reported the mean cosine similarity over each test set, multiplied by $100$. This evaluation encoder is separate from Falcon's audio encoder.

\paragraph{Ground-truth reference scores.}
Evaluating the ground-truth target speech on RealSSA-Sim yielded $11.83\%$ WER and $64.69$ speaker similarity under the same protocol. Since similarity was measured against an independent utterance from the same speaker, the ground-truth score provided an empirical reference for interpreting the absolute values. Falcon achieved $43.05$, leaving a gap of $21.64$ points to this reference. RealSSA-Real has no isolated target waveforms, so corresponding ground-truth scores are unavailable.

\paragraph{Computational efficiency.}
We reported trainable parameters, multiply-accumulate operations (MACs), CPU and GPU forward-pass times, and memory usage. Each measurement used a $2$\,s, $16$\,kHz audio clip paired with $50$ mouth-crop frames at $25$\,fps and a batch size of one. The hardware was an NVIDIA RTX 3090 GPU and an Intel Xeon Gold 6226R CPU. CPU measurements used one execution thread. We discarded five warmup runs and reported the median of the following $50$ forward passes. GPU time was measured with CUDA events, excluding host-side overhead. Memory measurements reported system RAM during CPU inference and peak GPU memory during GPU inference under the same protocol. Falcon required $51.3$\,ms on the GPU and $1361.2$\,ms on the CPU for the $2$\,s input.

\paragraph{Lookahead and streaming schedule.}
At $25$\,fps, the visual stream buffers three future frames, corresponding to $120$\,ms of lookahead. The audio STFT uses a $256$-sample window and a $128$-sample hop, corresponding to $16$\,ms and $8$\,ms at $16$\,kHz. The audio and visual timelines are aligned, the output advances at the $8$\,ms hop, and the configuration uses no additional future-chunk buffer.

Dividing the measured forward-pass times by approximately $250$ hops per $2$\,s input gives an amortized cost of $0.21$\,ms per hop on the GPU and $5.44$\,ms per hop on the CPU. These values describe average computation amortized over a complete input clip. Per-hop execution overhead and end-to-end latency require measurement with the streaming implementation, including input availability, window alignment, and output scheduling.

\section{Causalization of Dolphin and AV-Mossformer2 Baselines}
\label{app:baseline_causal}

Dolphin~\cite{li2026efficient} and AV-Mossformer2~\cite{zhao2025clearervoice} were originally released as non-causal AVTSE models. To compare them against Falcon under a common causal protocol, we re-implemented both baselines with the minimal modifications required to remove future-context access at every layer, while preserving their parameter counts, block topologies, and training objectives. Both baselines used the common optimizer and learning-rate schedule described in Section~\ref{sec:training}. This appendix lists, for each model, the operations that needed rewriting; every module not listed is already causal in the released code and was kept unchanged.

Across both models we used four rewriting primitives \cite{Sang2025AFA}. \emph{Causal self-attention} adds a lower-triangular mask before the softmax. \emph{Causal convolution} replaces symmetric padding by left padding only, with the same kernel size and dilation. \emph{Causal pooling and interpolation} replaces full-axis adaptive average pooling and nearest-neighbor interpolation along time by their one-sided (past-only) variants. \emph{Streaming-causal layer normalization} replaces full-axis LayerNorm or GroupNorm along time by the frame-local $\operatorname{LN}_{\text{c}}$ of Section~\ref{sec:separator}. Frame-wise operators, including pointwise convolutions, element-wise gates, and LayerNorm on the feature axis, were kept as released.

\subsection{Dolphin Causalization}
\label{app:dolphin_causal}

\paragraph{Visual encoder and refinement.}
The DP-LipCoder visual encoder \cite{li2026efficient} was frozen during AVTSE training, as in the original paper. Inside it we added causal masks to the temporal axis of every dual-path attention block, while the spatial axes remain bidirectional, and we rewrote every temporal downsample and upsample as a causal convolution. The post-encoder visual refinement block uses dilated depthwise convolutions with symmetric padding, which we rewrote as causal convolutions.

\paragraph{Audio-Visual Fusion.}
The fusion module has three paths: a resize path that upsamples the visual feature to the audio frame rate by nearest-neighbor interpolation, a key-value path that multiplies a pointwise-projected audio feature by the upsampled visual feature, and an attention path that softmaxes a projected visual feature over its time axis and upsamples the result to the audio time axis. The two non-causal operations are the nearest-neighbor upsamplings, which we rewrote as one-sided, and the softmax over the visual time axis, which we rewrote as a causal softmax that masks visual frames later than the current audio frame. The element-wise sum of the three paths is unchanged.

\paragraph{U-shape Separator.}
The Separator alternates Global Attention and Local Attention blocks across encoder and decoder stages, joined by injection-based skip fusion and a global Bottleneck. Global Attention compresses the input by adaptive average pooling along time, applies relative-position multi-head self-attention, and restores the resolution by nearest-neighbor upsampling; we rewrote the pooling and the upsampling as one-sided, added a causal mask to the inner attention, and tightened the relative-position clip to non-positive offsets. Local Attention contains Heat Diffusion Attention, which as released applies a discrete cosine transform (DCT) along the full time axis, a per-frequency exponential decay, and an inverse DCT. We replaced the full-sequence DCT pair with a causal windowed DCT on a past-only window of length $W$, so that the heat-diffusion interpretation is restricted to a finite past and the transform pair becomes a causal convolution of effective kernel $W$. The surrounding depthwise convolutions and GroupNorm layers in this block were rewritten as causal convolutions and as $\operatorname{LN}_{\text{c}}$, respectively. The Bottleneck aggregates encoder-stage outputs by adaptive average pooling along time before applying multi-head self-attention and a feed-forward network; we rewrote the aggregation pool as one-sided, added a causal mask to the attention, and rewrote the feed-forward depthwise convolution as causal. The injection-based skip fusion modules align the global feature to each skip's resolution by nearest-neighbor interpolation along time, which we rewrote as one-sided; the projection sub-modules inside the injection module reuse Heat Diffusion Attention and inherit its causal windowed DCT rewrite.

\paragraph{Audio encoder and decoder.}
The strided 1D audio encoder and its transposed-convolution decoder were rewritten as causal convolutions along the time axis.

\subsection{AV-Mossformer2 Causalization}
\label{app:avmossformer2_causal}

\paragraph{Pre-separator audio-visual fusion.}
The visual encoder was frozen as in the original paper \cite{zhao2025clearervoice}. The pre-separator fusion layer upsamples the visual feature to the audio frame rate by nearest-neighbor interpolation, which we rewrote as one-sided; the pointwise projection and the element-wise gated sum that follow are unchanged.

\paragraph{MossFormer block.}
Each MossFormer block runs a quadratic intra-group attention with squared-ReLU scores in parallel with a linearized cross-group attention. The released code already exposes a causal switch: when enabled, it adds a strictly upper-triangular mask to the squared-ReLU score tensor inside each group, and replaces the bidirectional cross-group statistics by one-sided cumulative sums followed by a one-step shift that excludes the current group from its own summary. We enabled this switch on every MossFormer block, and no further modification was required to the joint attention path.

\paragraph{Gated FSMN.}
The Gated FSMN module is already causal in the release: its memory taps span only past frames, the surrounding pointwise convolutions have time receptive field $1$, and the cumulative LayerNorm computes statistics only over past frames. The block was therefore kept unchanged.

\section{Per-Scene and Per-Speaker-Count Breakdown on RealSSA-Sim}
\label{app:scene_speaker_breakdown}

This appendix complements the aggregate RealSSA-Sim column of Table~\ref{tab:main-results} by resolving separation quality along the two axes that RealSSA-Sim was designed to vary jointly, namely the acoustic scene type and the number of near-field interfering speakers. Table~\ref{tab:scene_speaker_grid} reports SI-SNRi in decibels for every causal baseline and for Falcon on each $(\text{scene},\,N_{\text{near}})$ cell of the RealSSA-Sim test split, where $N_{\text{near}}$ denotes the number of near-field interfering speakers observed by the microphone, excluding the target. The cells of the grid exactly match the near-field interfering-speaker ranges listed per scene in Table~\ref{tab:srav_config_full}, so that $N_{\text{near}}=1$ corresponds to a mixture with one near-field interferer in addition to the target. Unlike the aggregate numbers of Table~\ref{tab:main-results}, this breakdown exposes the interaction between room acoustics and near-field crowding that a single RealSSA-Sim score averages over.

\begin{table}[h]
\centering
\caption{Per-scene and per-speaker-count SI-SNRi (dB, $\uparrow$) on the RealSSA-Sim test split. $N_{\text{near}}$ denotes the number of near-field interfering speakers in the mixture, excluding the target, so that $N_{\text{near}}=1$ is the one-interferer case. Columns group the four RealSSA-Sim scene types of Table~\ref{tab:srav_config_full}. For each column, \textbf{bold} marks the best result and \underline{underline} marks the second-best.}
\label{tab:scene_speaker_grid}
\resizebox{\linewidth}{!}{
\begin{tabular}{l ccccc ccc cccc cccc}
\toprule
\multirow{2}{*}{Method} & \multicolumn{5}{c}{Small meeting} & \multicolumn{3}{c}{Medium indoor} & \multicolumn{4}{c}{Large public} & \multicolumn{4}{c}{Outdoor} \\
\cmidrule(lr){2-6} \cmidrule(lr){7-9} \cmidrule(lr){10-13} \cmidrule(lr){14-17}
 & $1$ & $2$ & $3$ & $4$ & $5$ & $1$ & $2$ & $3$ & $1$ & $2$ & $3$ & $4$ & $1$ & $2$ & $3$ & $4$ \\
\midrule
AV-ConvTasNet~\cite{zhang2021online}     & $-0.6$ & 0.1 & 0.4 & 0.7 & 0.7 & 1.6 & 1.5 & 1.5 & 1.3 & 1.4 & 1.6 & 1.5 & 0.9 & 1.5 & 2.0 & 1.9 \\
AV-DPRNN~\cite{tao2025seanet}            & 4.0 & 4.4 & 4.3 & 4.3 & 4.1 & 5.1 & 4.9 & 4.8 & 5.4 & 5.4 & 5.5 & 5.6 & 6.6 & 6.8 & 7.0 & 7.1 \\
AV-Sepformer~\cite{lin2023av}            & 4.0 & 4.4 & 4.3 & 4.4 & 4.2 & 4.9 & 4.8 & 4.7 & 5.4 & 5.4 & 5.6 & 5.6 & 7.1 & 7.3 & 7.5 & 7.4 \\
CTCNet~\cite{li2024audio}                & 4.0 & 4.4 & 4.4 & 4.5 & 4.3 & 5.0 & 4.9 & 4.8 & 5.3 & 5.3 & 5.5 & 5.7 & 6.6 & 6.9 & 7.1 & 7.3 \\
RTFSNet-12~\cite{pegg2024rtfsnet}        & 2.5 & 3.4 & 3.5 & 3.9 & 3.7 & 4.5 & 4.6 & 4.5 & 4.7 & 4.9 & 4.9 & 5.3 & 4.8 & 5.4 & 5.8 & 6.0 \\
AV-Mossformer2~\cite{zhao2025clearervoice} & 4.6 & 5.0 & 5.0 & 5.0 & 4.7 & 5.4 & 5.3 & 5.1 & 5.8 & 5.8 & 6.0 & 6.1 & 7.7 & 8.0 & 7.9 & 8.0 \\
Dolphin~\cite{li2026efficient}           & 5.1 & 5.6 & 5.5 & 5.5 & 5.3 & 6.3 & 6.2 & 6.0 & 6.8 & 7.0 & 7.1 & 7.3 & 8.7 & 9.1 & 9.2 & 9.4 \\
AV-TF-GridNet~\cite{pan2023scenario}     & \underline{6.4} & \underline{6.8} & \underline{7.0} & \underline{7.0} & \textbf{6.9} & \underline{7.1} & \underline{7.0} & \underline{6.7} & \underline{7.5} & \textbf{7.6} & \textbf{7.8} & \textbf{8.0} & \underline{8.8} & \underline{9.1} & \textbf{9.4} & \underline{9.4} \\
Swift-Net-12~\cite{Sang2025AFA}          & 6.1 & 6.5 & 6.6 & 6.6 & 6.3 & 6.9 & 6.9 & 6.7 & 7.2 & 7.3 & 7.3 & 7.5 & 8.5 & 8.7 & 8.8 & 8.8 \\
\midrule
\textbf{Falcon (Ours)} & \textbf{6.7} & \textbf{7.0} & \textbf{7.1} & \textbf{7.1} & \underline{6.8} & \textbf{7.2} & \textbf{7.2} & \textbf{7.1} & \textbf{7.6} & \underline{7.6} & \underline{7.8} & \underline{7.8} & \textbf{9.0} & \textbf{9.1} & \underline{9.3} & \textbf{9.4} \\
\bottomrule
\end{tabular}}
\end{table}

Three patterns emerge from the grid. The scene axis did not follow a simple monotone difficulty ordering with respect to reverberation time. Outdoor cells consistently yielded the highest SI-SNRi for every architecture because the near-anechoic propagation of Table~\ref{tab:srav_config_full} leaves the target direct path largely intact, so that the separator only needs to suppress wide-band ambient noise and a small number of geometrically well separated interferers. Large public cells were the next easiest on average, since the dominant interference is reverberant rather than spatially co-located with the target and the visual enrollment cue is sufficient to disambiguate far-field talkers. Medium indoor cells lay in the middle and combine moderate reverberation with the first far-field interferers. Small meeting cells were the hardest, because the near-field interferers lie inside the critical distance of the microphone and therefore inject energy whose spectral envelope, onset timing, and reverberation fingerprint closely match those of the target. The fact that a low-reverberation outdoor scene produces higher SI-SNRi than a short-RT60 small meeting confirms that crowding of near-field talkers, rather than reverberation alone, is the dominant difficulty axis on RealSSA-Sim.

The speaker-count axis followed a non-monotone trajectory that is characteristic of audio-visual target speaker extraction rather than of blind speech separation. In the outdoor scene every competitive model saw its SI-SNRi rise as $N_{\text{near}}$ grows from one to four, for instance Falcon improved from $9.0$\,dB at $N_{\text{near}}=1$ to $9.4$\,dB at $N_{\text{near}}=4$, which is consistent with the input mixture becoming progressively more distorted and therefore admitting a larger improvement margin before the metric saturates. In the small meeting scene the improvement peaked around three or four near-field speakers and then declined, with Falcon rising from $6.7$\,dB at $N_{\text{near}}=1$ to $7.1$\,dB at $N_{\text{near}}=3$ and $N_{\text{near}}=4$ before dropping to $6.8$\,dB at $N_{\text{near}}=5$. This non-monotonicity arises because once a small room contains four or more nearby talkers with closely matched reverberation fingerprints, the visual enrollment cue remains able to resolve the target identity, but the total near-field interfering energy begins to exceed the budget that a single-channel mask can suppress while still preserving the target spectrum. The aggregate RealSSA-Sim column of Table~\ref{tab:main-results} averages over this behavior and therefore underreports the cell-level gap that separates strong causal architectures from lightweight ones in the crowded-room regime.

The grid further shows that Falcon extended the quality of heavyweight causal baselines to conditions in which lightweight ones collapsed. Falcon obtained the best SI-SNRi on eleven of the sixteen cells and the second-best on five more. In the small-meeting and medium-indoor cells, its SI-SNRi margin over AV-TF-GridNet ranged from $-0.1$ to $+0.4$\,dB despite using $91.9\%$ fewer parameters (Table~\ref{tab:efficiency}); its margin over Swift-Net-12 was $0.3$ to $0.6$\,dB at the same parameter count. The lightweight iterative baselines lost ground most sharply on this same column, with RTFSNet-12 falling below $4$\,dB on every small meeting cell and AV-ConvTasNet dropping below $1$\,dB, which indicates that repeated application of a small separator cannot by itself compensate for the spectrally similar near-field interference that small rooms produce. The single-pass multi-scale separator and the encoder-space speaker contrastive loss of Falcon therefore act where they were designed to, namely in the crowded near-field cells that the aggregate RealSSA-Sim score tends to wash out, while remaining within $0.2$\,dB of the strongest heavyweight baseline in the reverberation-dominated large public and outdoor cells.

\section{Qualitative Target Speech Extraction Results}
\label{app:qualitative}

\begin{figure}[h]
  \centering
  \begin{subfigure}{0.95\linewidth}
    \centering
    \includegraphics[width=\linewidth]{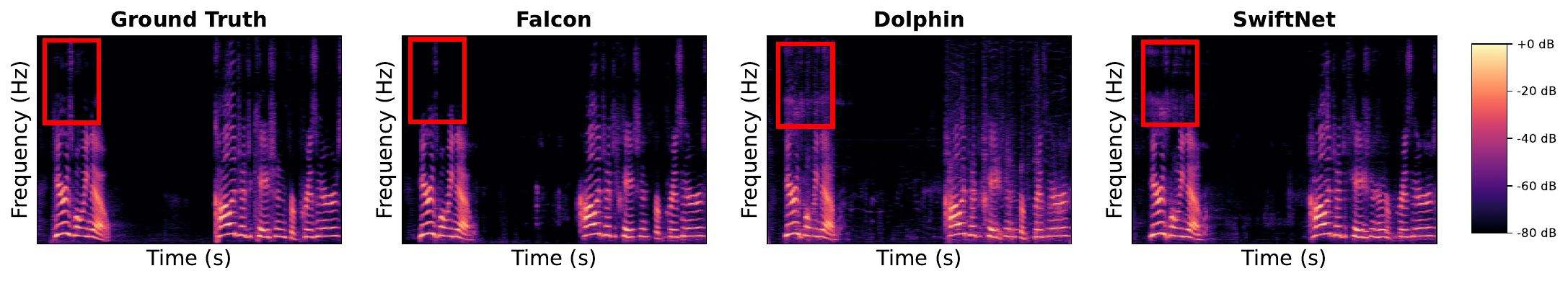}
    \caption{Demo I}
    \label{fig:qual-demo1}
  \end{subfigure}

  \begin{subfigure}{0.95\linewidth}
    \centering
    \includegraphics[width=\linewidth]{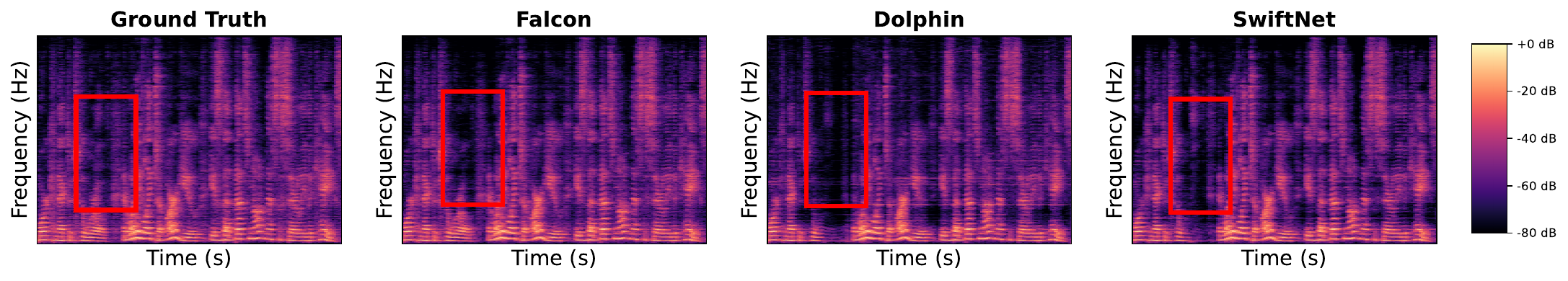}
    \caption{Demo II}
    \label{fig:qual-demo2}
  \end{subfigure}

  \begin{subfigure}{0.95\linewidth}
    \centering
    \includegraphics[width=\linewidth]{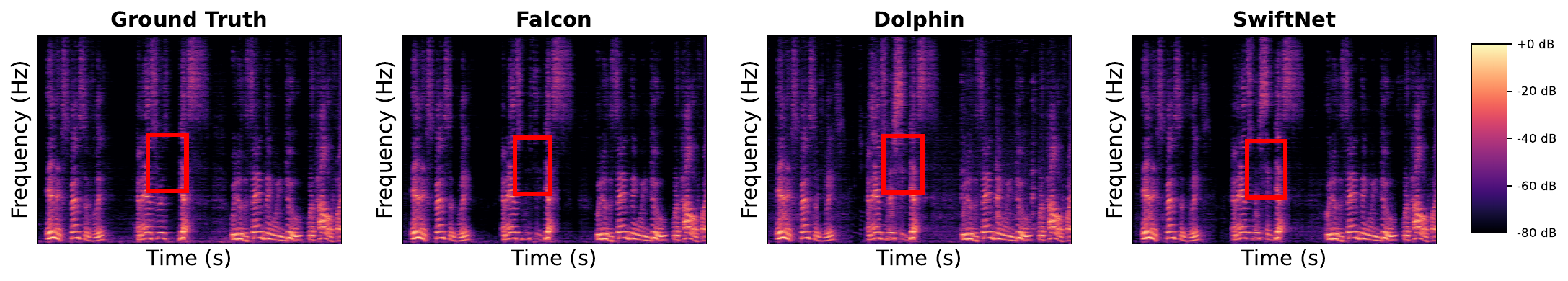}
    \caption{Demo III}
    \label{fig:qual-demo3}
  \end{subfigure}

  \begin{subfigure}{0.95\linewidth}
    \centering
    \includegraphics[width=\linewidth]{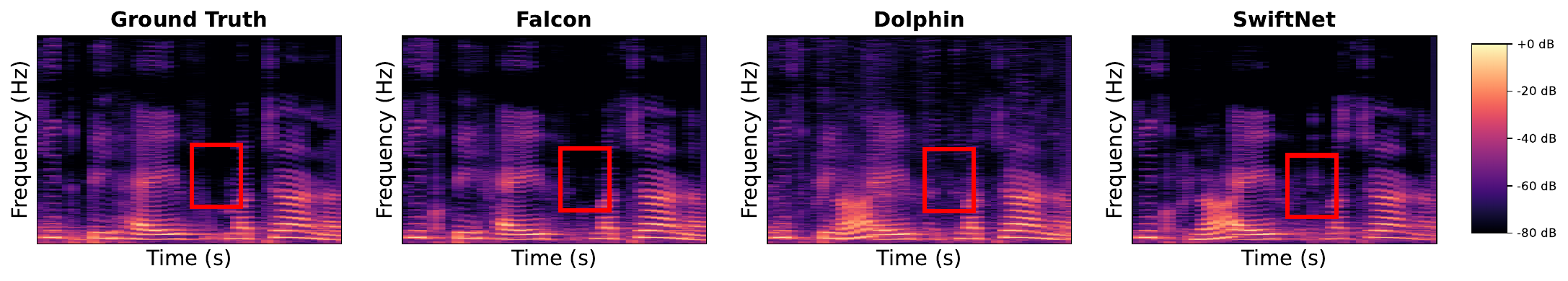}
    \caption{Demo IV}
    \label{fig:qual-demo4}
  \end{subfigure}

  \caption{Qualitative spectrogram comparison of target speech extraction results on four representative test mixtures. Each example compares the ground-truth target speech with the outputs of Falcon, Dolphin, and SwiftNet. Red boxes highlight regions where the methods differ in preserving target speech components or suppressing residual interference.}
  \label{fig:qualitative-spectrogram}
\end{figure}

We conducted a qualitative spectrogram analysis of target speech extraction results. The models were evaluated on representative test mixtures, and the results are shown in Figure~\ref{fig:qualitative-spectrogram}. For each example, we compared the ground-truth target speech with the outputs of Falcon, Dolphin, and SwiftNet. The highlighted regions mark representative time-frequency areas where the methods differ in preserving target speech components and suppressing residual interference.

In Figure~\ref{fig:qualitative-spectrogram}\subref{fig:qual-demo1}, the highlighted region contains a target speech component that is clearly visible in the ground-truth spectrogram. Falcon preserved this component with a structure closer to the ground truth, while keeping the surrounding non-target regions relatively suppressed. Dolphin retained more diffuse residual energy, and SwiftNet showed stronger residual interference around the highlighted region.

In Figure~\ref{fig:qualitative-spectrogram}\subref{fig:qual-demo2}, the mixture contains dense time-frequency patterns, making target extraction more challenging. Falcon better matched the ground-truth structure in the highlighted region and suppressed surrounding interference more effectively. By contrast, Dolphin and SwiftNet left more residual energy, especially around the low-frequency and neighboring time regions.

In Figure~\ref{fig:qualitative-spectrogram}\subref{fig:qual-demo3}, the target speech component is relatively narrow and localized. Falcon preserved this localized structure more faithfully and maintained cleaner background regions. Dolphin and SwiftNet recovered the main component, but their outputs contained more residual energy from nearby speech components, indicating weaker suppression of interfering speakers.

In Figure~\ref{fig:qualitative-spectrogram}\subref{fig:qual-demo4}, the target speech shows a clear harmonic pattern. Falcon reconstructed a more coherent harmonic structure and better suppressed non-target components. Dolphin produced a more diffuse spectrogram with noticeable residual interference, while SwiftNet tended to over-suppress parts of the speech structure and still left residual artifacts.

Across all four examples, Falcon produced spectrograms that were closer to the ground-truth target speech, with clearer target structures and less residual interference than the compared baselines.

\section{Recurrent Unit inside TFRB}
\label{app:tfrb_rnn}

Each TFRB operates both at the bottleneck and at the decoder levels (Section~\ref{sec:separator}), so its recurrent unit must retain long-range context at low resolutions and preserve fine spectral structure after fusing multi-scale skip features. Table~\ref{tab:rnn-ablation} compares four candidates. LSTM achieved the highest SI-SNRi ($7.7$\,dB), exceeding GRU by $0.4$\,dB at the cost of $0.05$\,M parameters and $1.27$\,G MACs; its explicit cell state decouples long-term retention from short-term gating, which helps when the same module spans resolutions differing by factors of two to four. SRU matched GRU in quality with fewer parameters, but neither closed the gap to LSTM.

\begin{table}[h]
  \caption{Recurrent unit comparison in TFRB. LSTM achieved the highest SI-SNRi with moderate parameter and computational overhead.}
  \label{tab:rnn-ablation}
  \centering
  \begin{tabular}{lcccc}
    \toprule
    Recurrent Unit & SI-SNRi $\uparrow$ & SDRi $\uparrow$ & Params (M) $\downarrow$ & MACs (G) $\downarrow$ \\
    \midrule
    RNN            & 6.2   & 7.1   & 0.08 & 1.90 \\
    SRU            & 7.4   & 7.9   & 0.15 & 2.71 \\
    GRU            & 7.3   & 7.9   & 0.20 & 4.42 \\
    \midrule
    \textbf{LSTM (Ours)} & \textbf{7.7} & \textbf{8.2} & 0.25 & 5.69 \\
    \bottomrule
  \end{tabular}
\end{table}

\section{Visualization of Encoder-Space Speaker Embeddings}
\label{app:loss-visualization}

To further analyze the effect of the proposed encoder-space speaker contrastive loss, we visualized speaker embeddings extracted from models trained with and without this loss. We selected the top-15 target speakers with the most utterances from the RealSSA-Sim training set, resulting in 5--8 utterances per speaker. For each utterance, we extracted an encoder-space speaker embedding using the speaker pooling head and projected all embeddings into two dimensions using principal component analysis (PCA).

\begin{figure}[h]
  \centering
  \includegraphics[width=0.95\linewidth]{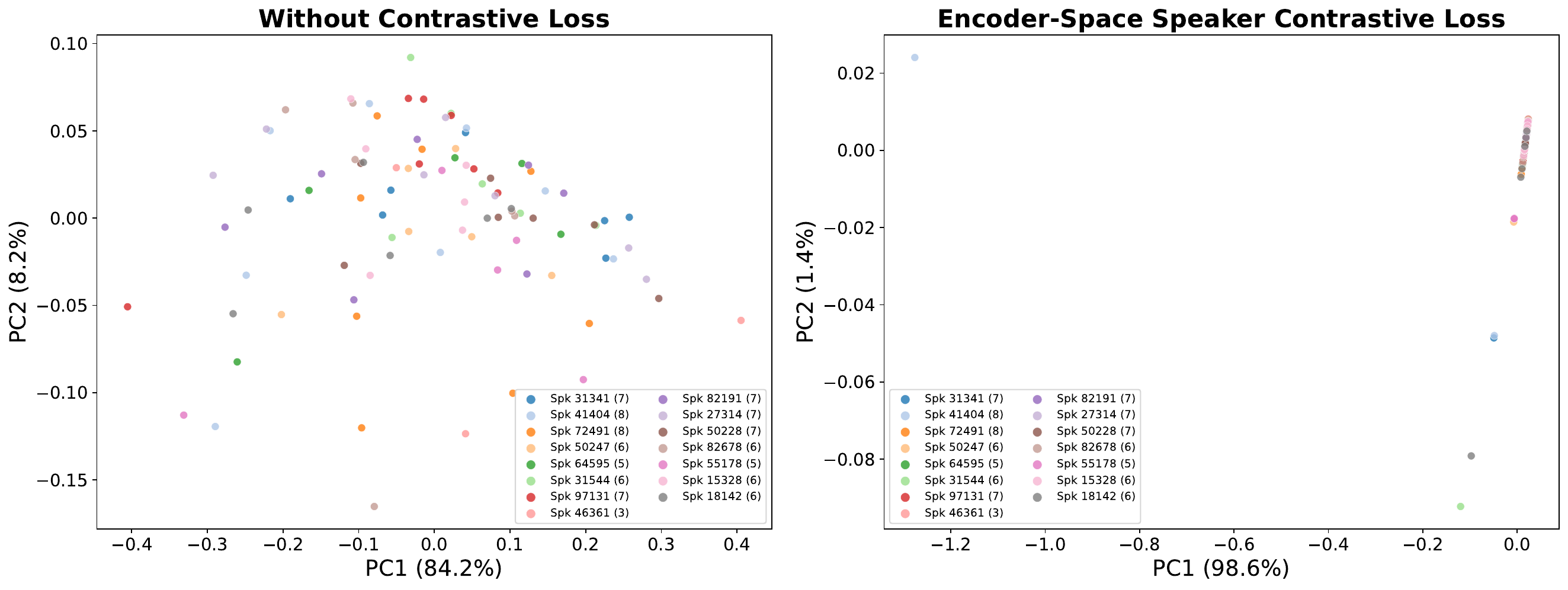}
  \caption{PCA visualization of encoder-space speaker embeddings for the top-15 target speakers in the RealSSA-Sim training set. Left: without the proposed contrastive loss, embeddings from different speakers are scattered and overlap. Right: with the contrastive loss, same-speaker embeddings become more compact and different speakers are more clearly separated.}
  \label{fig:loss-pca}
\end{figure}

As shown in Figure~\ref{fig:loss-pca}, without the contrastive loss, embeddings from different speakers were scattered and overlapped with each other, indicating that the encoder-space representations were weakly structured with respect to speaker identity. In contrast, with the proposed contrastive loss, same-speaker embeddings became more compact and different speakers were more clearly separated. This visualization supports that the proposed loss introduces a speaker-discriminative structure in the encoder space, which helps the separator distinguish the target speaker from near-field interferers.

\section{Output Formulation Comparison}
\label{app:output_formulation}

We compared a complex-valued mask, a real-valued mask, and direct prediction of real and imaginary components on RealSSA-Sim. The backbone, training configuration, and evaluation protocol were kept the same across the three output formulations. Table~\ref{tab:output_formulation} reports the results.

\begin{table}[t]
\centering
\caption{Comparison of output formulations on RealSSA-Sim. SI-SNRi and SDRi are reported in dB; higher is better. Bold indicates the best result.}
\label{tab:output_formulation}
\begin{tabular}{lcc}
\toprule
Output formulation & SI-SNRi $\uparrow$ & SDRi $\uparrow$ \\
\midrule
Complex-valued mask (Falcon) & \textbf{7.7} & \textbf{8.2} \\
Real-valued mask & 7.3 & 7.9 \\
Direct real/imaginary prediction & 7.1 & 7.2 \\
\bottomrule
\end{tabular}
\end{table}

The complex-valued mask improved SI-SNRi and SDRi by $0.4$\,dB and $0.3$\,dB over the real-valued mask, and by $0.6$\,dB and $1.0$\,dB over direct real/imaginary prediction. These results support the complex-valued mask formulation used by Falcon under the evaluated model and training configuration.

\section{Performance across Speech Overlap Ratios}
\label{app:overlap_ratio}

We grouped all $9{,}169$ RealSSA-Sim test samples by speech overlap ratio to examine extraction performance at different degrees of simultaneous speech. A higher overlap ratio indicates that multiple speakers are active simultaneously for a larger fraction of the utterance. Table~\ref{tab:overlap_ratio} gives the sample count, mean overlap ratio, and Falcon's extraction performance for each group.

\begin{table}[t]
\centering
\caption{Falcon's performance across speech overlap ratios on RealSSA-Sim. SI-SNRi and SDRi are reported in dB.}
\label{tab:overlap_ratio}
\begin{tabular}{ccccc}
\toprule
Overlap range & Samples & Mean overlap & SI-SNRi $\uparrow$ & SDRi $\uparrow$ \\
\midrule
$0.11$--$0.72$ & $1{,}834$ & $0.635$ & $7.95$ & $8.39$ \\
$0.72$--$0.79$ & $1{,}834$ & $0.756$ & $7.71$ & $8.30$ \\
$0.79$--$0.84$ & $1{,}834$ & $0.811$ & $7.80$ & $8.33$ \\
$0.84$--$0.89$ & $1{,}834$ & $0.863$ & $7.73$ & $8.28$ \\
$0.89$--$1.00$ & $1{,}833$ & $0.946$ & $7.73$ & $8.22$ \\
\bottomrule
\end{tabular}
\end{table}

Across the five groups, Falcon's SI-SNRi ranged from $7.71$ to $7.95$\,dB and SDRi from $8.22$ to $8.39$\,dB. In the highest-overlap group, whose mean overlap ratio is $0.946$, Falcon achieved $7.73$\,dB SI-SNRi and $8.22$\,dB SDRi. These scores were $0.22$\,dB and $0.17$\,dB below the lowest-overlap group, respectively, showing similar extraction improvements across the reported groups.

\section{Missing Visual Information}
\label{app:visual_missing}

We evaluated sensitivity to missing visual information on a fixed random subset of $500$ RealSSA-Sim test samples. The missing ratio $p$ was set to $0$, $0.25$, $0.50$, $0.75$, or $1.00$. Random Drop removes visual frames at random to simulate intermittent information loss, whereas Block Drop masks a continuous segment to simulate sustained information loss. All methods used the same samples and missing-information settings.

\begin{table}[t]
\centering
\caption{Performance under missing visual information on $500$ RealSSA-Sim test samples. Each entry reports SI-SNRi / SDRi in dB. The $p=0$ column is the uncorrupted control for this subset. At $p=1$, the two missing patterns are equivalent. Bold indicates the best result for each metric under the same pattern and missing ratio.}
\label{tab:visual_missing}
\resizebox{\linewidth}{!}{%
\begin{tabular}{llccccc}
\toprule
Method & Missing pattern & $p=0$ & $p=0.25$ & $p=0.50$ & $p=0.75$ & $p=1.00$ \\
\midrule
AV-Sepformer~\cite{lin2023av} & Random Drop & 5.26 / 5.88 & 4.98 / 5.68 & 4.57 / 5.38 & 2.35 / 4.02 & $-44.09$ / $-4.66$ \\
AV-Sepformer~\cite{lin2023av} & Block Drop & 5.26 / 5.88 & 2.42 / 3.65 & $-4.17$ / $-0.58$ & $-12.42$ / $-3.46$ & $-44.09$ / $-4.66$ \\
RTFSNet~\cite{pegg2024rtfsnet} & Random Drop & 4.01 / 4.61 & 3.87 / 4.50 & 3.42 / 4.15 & 1.69 / 3.02 & $-10.80$ / $-1.73$ \\
RTFSNet~\cite{pegg2024rtfsnet} & Block Drop & 4.01 / 4.61 & 2.77 / 3.66 & 0.51 / 2.20 & $-3.38$ / 0.31 & $-10.80$ / $-1.73$ \\
CTCNet~\cite{li2024audio} & Random Drop & 5.07 / 5.72 & 4.83 / 5.53 & 4.30 / 5.09 & 2.90 / 4.03 & $-14.85$ / $-3.72$ \\
CTCNet~\cite{li2024audio} & Block Drop & 5.07 / 5.72 & 3.04 / 4.10 & $-0.28$ / 1.76 & $-4.01$ / $-0.39$ & $-14.85$ / $-3.72$ \\
\midrule
\textbf{Falcon} & Random Drop & \textbf{7.48 / 7.88} & \textbf{7.28 / 7.72} & \textbf{6.75 / 7.30} & \textbf{5.97 / 6.73} & \textbf{$-9.91$ / 0.13} \\
\textbf{Falcon} & Block Drop & \textbf{7.48 / 7.88} & \textbf{6.23 / 6.84} & \textbf{3.46 / 4.78} & \textbf{0.01 / 2.74} & \textbf{$-9.91$ / 0.13} \\
\bottomrule
\end{tabular}%
}
\end{table}

Falcon achieved the highest SI-SNRi and SDRi in all evaluated partial-missing conditions. Under Random Drop, its performance decreased gradually as the missing ratio increased, retaining $5.97$\,dB SI-SNRi when $75\%$ of the visual frames were missing. At the same ratio, Block Drop reduced SI-SNRi to $0.01$\,dB. Sustained visual loss therefore had a larger effect on extraction performance than dispersed frame loss in this evaluation.

When visual information was entirely absent, all compared methods degraded, and Falcon's SI-SNRi fell to $-9.91$\,dB. The current systems depend on visual information to identify the target speaker. These experiments characterize intermittent and sustained loss of visual information; sensitivity to mouth-crop displacement and tracking a different speaker remains to be evaluated separately.

\section{LLM Usage Statement}
\label{app:llm_usage}

We used large language models as general-purpose writing assistants for polishing sentence-level English phrasing in the manuscript. All scientific contributions, including the definition of the RealSSA benchmark, the design of the Separator and the encoder-space speaker contrastive loss, the experimental protocol, and the analysis of results, were conceived, implemented, verified, and written by the authors. No LLM was used to generate experimental results, to draft novel technical claims, or to position the work against related literature. The authors take full responsibility for the content of the paper, including any residual errors.

\end{document}